\documentclass{raa04}

\renewcommand{\tablename}{\bf{Table}}
\makeatletter

\newcommand{\Rmnum}[1]{\expandafter\@slowromancap\romannumeral #1@}
\makeatother

\def\cm2{cm$^{-2}$}

\def\nh3{NH$_3$}
\def\n2h{N$_2$H$^+$}

\def\13co{$^{13}$CO}
\def\c18o{C$^{18}$O}
\def\ch3cch{CH$_3$CCH}
\def\h2o{H$_2$O}
\def\o2{O$_2$}
\def\hc3n{HC$_3$N}
\def\h2{H$_2$}
\def\nh{n(H$_2$)}

\def\616-523{6$_16$-5$_23$}
\def\110-101{1$_10$-1$_01$}
\def\h216o{H$_2$$^{16}$O}
\def\r0{R$_0$}
\def\hii{H{\footnotesize\Rmnum{2}\ }}
\usepackage{graphicx,times}
\usepackage{amssymb,amsmath}
\usepackage{multirow}

\usepackage{natbib}
\usepackage{upgreek}
\bibpunct{(}{)}{;}{a}{}{,}

\begin{document}
   \title{Water abundance in four of the brightest water sources in the southern sky
}

\volnopage{{\bf 2016} Vol.\ {\bf 16} No. {\bf 3}, ~ 39 (16pp)~
 {\small  doi: 10.1088/1674--4527/16/3/039}}
      \setcounter{page}{1}

   \author{Bing-Ru Wang\inst{1}, Lei Qian\inst{1}, Di Li\inst{1,2} \and Zhi-Chen Pan\inst{1}}

   \institute{ National Astronomical Observatories,
Chinese Academy of Sciences, Beijing 100012, China; {\it brwang@nao.cas.cn}\\
        \and
             Key Laboratory of Radio Astronomy, Chinese Academy of Sciences,
             Nanjing 210008, China\\
\vs\vs \no
   {\small Received 2015 April 3; accepted 2015 September 11}
}

\abstract{  We estimated the ortho-{\rm{H$_2$O}} abundances of
G267.9--1.1, G268.4--0.9, G333.1--0.4 and G336.5--1.5, four of the
brightest ortho-{\rm{H$_2$O}} sources in the southern sky observed
by the Submillimeter Wave Astronomy Satellite (ortho-{\rm{H$_2$O}}
1$_{10}$ -- 1$_{01}$ line, 556.936~GHz). The typical molecular
clumps in our sample have H$_2$ column densities of $10 ^{22}$ to
$10 ^{23}${\,}cm$^{-2}$ and ortho-{\rm{H$_2$O}} abundances of
10$^{-10}$. Compared with previous studies, the ortho-{\rm{H$_2$O}}
abundances are at a low level, which can be caused by the low
temperatures of these clumps. To estimate the ortho-{\rm{H$_2$O}}
abundances, we used the CS $J = 2 \to  1$ line (97.98095~GHz) and CS
$J = 5 \to  4$ (244.93556~GHz) line observed by{ the} Swedish-ESO 15\,m
Submillimeter Telescope (SEST) to calculate the temperatures of the
clumps and the 350~$\upmu$m dust continuum observed by{ the} Caltech
Submillimeter Observatory (CSO) telescope to estimate the H$_2$
column densities. The observations of {\rm{N$_2$H$^+$}} ($J = 1 \to
0$) for these clumps were also acquired by SEST and the
corresponding abundances were estimated. The {\rm{N$_2$H$^+$}}
abundance in each clump shows a common decreasing trend toward the
center and {a} typical abundance range from
10$^{-11}$ to 10$^{-9}$. 
}

   \authorrunning{B.-R. Wang et al. }
   \titlerunning{Water Abundance in Four of the Brightest
    Water Sources}
   \maketitle

\section{Introduction}           
\label{sect:intro}

Water was first detected in the interstellar medium{ (ISM)} over 40 years
ago (\citealt{Cheung+etal+1969}). It is an essential coolant in
star-forming regions and plays an important role in the energy
balance of prestellar objects (\citealt{Doty+Neufeld+1997}). Thus,
the abundance of water is a crucial parameter, especially for
massive star formation (\citealt{Emprechtinger+etal+2010}). Since
the physical conditions of star-forming regions affect the water
abundance (with respect to H$_2$), water {acts as} an excellent diagnostic {for} energetic phenomena
(\citealt{Kristensen+vanDishoeck+2011}). As an abundant
oxygen-bearing molecule formed in molecular clouds, its abundance
also gives constraints {on} the abundance of atomic oxygen, therefore
{it }affects the abundances of other chemically related oxygen-bearing
species.

Accessible water lines and feasible methods are necessary for
estimating the abundance of water in star-forming-regions. Water
lines originating from different levels probe gas under different
conditions. Most rotational water lines, including the ground-state
transition of ortho- and para-{\rm{H$_2$O}}, cannot be observed
{from} the ground due to the existence of telluric water
(\citealt{Emprechtinger+etal+2010}). Although there {are} indeed
some transitions {that have been }detected {from} the ground, their
upper states are over 200~K above the ground state
(\citealt{Snell+etal+2000a}). The high energies over the ground
state indicate high gas temperatures when collision with H$_2$ is
considered as the excitation mechanism. Thus, these transitions are
unlikely{ to be} from cold gases. To date, space observations,
({\rm{e.g.}}, the Submillimeter Wave Astronomy Satellite (SWAS)
(\citealt{Melnick+etal+2000}); the Odin satellite; the Infrared
Space Observatory {(}ISO{)}; the Spitzer Space Telescope and the
Herschel Space Observatory) have detected water lines, including the
556.936~GHz ortho-{\rm{H$_2$O}} 1$_{10}$ -- 1$_{01}$ line. This
ground-state transition was observed by SWAS first. With the upper
state lying only 27~K above the ortho-{\rm{H$_2$O}} ground state, it
provides access to estimate the water abundance in cold molecular
gas, in which massive stars form in cold dense clumps and {young
stellar objects} are deeply buried.

Water can form in several different routes, both in {gas phase} and
on dust grains. Once they form, the {\rm{H$_2$O}} molecules can be
desorbed from the ice mantle of dusts, remain frozen on the dust
surface or freeze onto the dust grains from the gas phase. Water ice
on the dust surface can desorb thermally when the dust temperature
{rises} above about 100~K (\citealt{Hollenbach+etal+2009}). In
another way, photodesorption occurs when the ice absorbs {ultraviolet
(}UV{)} photons (\citealt{vanDishoeck+etal+2013}). When the
temperature is as low as about 10\,K and the density is high enough,
freeze-out will dominate (\citealt{Bergin+vanDishoeck+2012}) {and
}consequently lead to low {\rm{H$_2$O}} abundances. Thus, temperature
and UV radiation are essential {factors} that affect water abundance.

To compare observations {with} predicted results, the abundances
of para- or ortho-{\rm{H$_2$O}} are estimated based on the spectra
obtained from telescopes. For the ortho-{\rm{H$_2$O}} line
(1$_{10}$ -- 1$_{01}$, 556.936~GHz), an effectively optically thin
approximation (\citealt{Snell+etal+2000a}) was adopted, which
makes it convenient to estimate the ortho-{\rm{H$_2$O}} abundance.
In the Herschel key programme ``water in star-forming regions with
Hersehel" (WISH), the 
non-local thermodynamic equilibrium (LTE) radiative
  transfer
code RADEX (\citealt{vanderTak+etal+2007}) was used to reduce the
ortho-{\rm{H$_2$O}} line (1$_{10}$ -- 1$_{01}$, 556.936~GHz) data
to estimate the {\rm{H$_2$O} abundance in{ a} low-mass protostar
(\citealt{Kristensen+etal+2012}). In this paper, we estimate the
ortho-{\rm{H$_2$O}} abundances of four of the brightest
ortho-{\rm{H$_2$O}} sources (G267.9--1.1, G268.4--0.9, G333.1--0.4
and G336.5--1.5) in the southern sky observed by SWAS. The paper
is organized as {follows}: in Section 2, we briefly describe the
observations of these clumps and the data reduction procedures. In
Section 3, we present the calculations and estimates of clump
temperatures, clump masses, H$_2$ column densities and finally the
estimate of ortho-{\rm{H$_2$O}} abundances based on observations.
In Section 4 and Section 5, we present the discussion and
conclusion respectively. The appendix contains some supplementary
material.

\section{Observation and Data Reduction}
\label{sect:Obs}
\subsection{Source Selection}

We checked the co-added spectra of{ the} ortho-{\rm{H$_2$O}} 1$_{10}$ --
1$_{01}$ line of all the 386 sources in the five and a half years
of the SWAS nominal mission from Lambda\footnote{\it
http://lambda.gsfc.nasa.gov/product/swas/s\_sw.cfm}. We selected
four of the sources with $T_{\rm{A}}^*$ higher than 0.1~K
(exclud{ing} the sources in the Galactic Center region) in the
southern sky. These four sources are G267.9--1.1 ($T_{\rm{A}}^*$ =
0.10~K), G268.4--0.9 ($T_{\rm{A}}^*$ = 0.16~K), G333.1--0.4
($T_{\rm{A}}^*$ = 0.20~K) and G336.5--1.5 ($T_{\rm{A}}^*$ = 0.45\
K). Among these four sources, G336.5--1.5 has the highest
$T_{\rm{A}}^*$.

These four sources {are located} in star forming
regions RCW 38 (G267.9--1.1 and G268.4--0.9), RCW 106 (G333.1--0.4)
and RCW 108 (G336.5--1.5), respectively. Although being bright at
8~$\upmu$m, they are all associated with the 22~GHz 6$_{16}$
--5$_{23}$ water masers (\citealt{Kaufmann+etal+1976};
\citealt{Braz+etal+1989}; \citealt{Caswell+etal+1974} and
\citealt{Valdettaro+etal+2007}), which are believed to be good
indicators of the location of massive star formation
(\citealt{Juvela+1996}). The properties of these four sources are
summarized briefly as {follows}.

\begin{itemize}
  \item[(1)] G267.9--1.1.  It is the third brightest source in the
investigation of Galactic radio sources at 5000~MHz
(\citealt{Goss+Shaver+1970}), with a brightness
 temperature of
124.0~K (\citealt{Shaver+Goss+1970}). The associated 22~GHz
6$_{16}$ --5$_{23}$ water maser (without OH main-line emission)
was first reported by \cite{Kaufmann+etal+1976}.

\item[(2)] G268.4--0.9.  It was identified in an 11~cm survey of Vela
(\citealt{Manchester+Goss+1969}), near the source
G267.9--1.1{ }(denoted as G268.0--1.0 in the same survey) with a
lower brightness temperature. However, it was not identified as an
isolated radio source in the Galactic radio source survey
(\citealt{Goss+Shaver+1970}). The associated 22~GHz water maser
was identified by \cite{Braz+etal+1989}.

\item[(3)] G333.1--0.4.  It is one of the clumps in the giant molecular
cloud (GMC) G333 (\citealt{Lowe+etal+2014}). It was first
identified as an extensive \hii region (\citealt{Beard+1966}),
with a brightness temperature of 17.8~K
(\citealt{Shaver+Goss+1970}). The associated 22~GHz water maser
was identified by \cite{Caswell+etal+1974}. The
ortho-{\rm{H$_2$O}} 556.936~GHz line obtained by the SWAS exhibits
a pronounced inverse P{ }Cygni profile and related study
(\citealt{Li+etal+2004}) suggests that it is a rare case of direct
observational evidence for large scale infall in a star forming
region.

\item[(4)] G336.5--1.5.  It is identified as an isolated compact \hii
region in both the survey of H109$\alpha$ recombination line
emission in Galactic \hii regions of the southern sky
(\citealt{Wilson+etal+1970}) and the investigation of Galactic
radio sources at 5000 MHz (\citealt{Goss+Shaver+1970}). Its
brightness temperature is 7.2~K (\citealt{Shaver+Goss+1970}).
G336.5--1.5 is associated with {bright-rimmed cloud (}BRC{)} 79,
one of the 89 clouds in a catalog of {BRCs} with IRAS point
sources (\citealt{Sugitani+Ogura+1994}). It has the largest H$_2$
column density among the 43 southern hemisphere BRCs (BRC 77 and
BRC 78 excluded) and the \hii region RCW 62, according to
$^{13}$CO observations (\citealt{Yamaguchi+etal+1999}). Its
ortho-{\rm{H$_2$O}} 556.936~GHz line obtained by SWAS exhibits the
highest antenna temperature among all observed sources (other than
solar system objects and the Galactic Center), which makes it an
interesting object to study.
\end{itemize}

Compared with {the }other {three} sources, G336.5--1.5 has a
higher Galactic latitude. Its associated 22~GHz water maser was
detected in a survey of 45 southern BRCs
(\citealt{Sugitani+Ogura+1994}) for H$_2$O maser emission
(\citealt{Valdettaro+etal+2007}), with a total integrated H$_2$O
flux density of merely 5.4~Jy~km~s$^{-1}$. All these features
mentioned above imply that these four sources are likely to be
massive star forming active clumps. We use ``clumps" to refer to
these four sources in this paper.

\subsection{Observation and Data Reduction}

The observations were carried out with three telescopes. The
ortho-{\rm{H$_2$O}} 1$_{10}$ -- 1$_{01}$ line (556.936~GHz) was
observed with SWAS. The CS $J = 2 \to 1$  line (97.98095~GHz), CS $J
= 5 \to  4$ line (244.93556~GHz) and {\rm{N$_2$H$^+$}} $J = 1 \to  0$
line (93.17340~GHz) data {were} from the Swedish-ESO 15\,m
Submillimeter Telescope (SEST\footnote{\it
http://www.eso.org/public/images/esopia00049teles/}). The
350~$\upmu$m dust continuum data {were} obtained with the
Submillimeter High Angular Resolution Camera~II (SHARC II; see
\citealt{Dowell+etal+2003}) of the Caltech Submillimeter Observatory
(CSO) telescope. The observational parameters of molecular lines
{are} listed in Table~\ref{observation}.

\begin{table*}
\centering

\begin{minipage}{70mm}
\caption{Observational Parameters of Molecular Lines
\label{observation}}\end{minipage}

\renewcommand\baselinestretch{1.4}
 \fns \tabcolsep 3mm
 \begin{tabular}{lccccc}
  \hline\noalign{\smallskip}
Transition & Frequency & Instrument & Beam Size & $\Delta \nu $ & $\Delta v$ \\
           & (GHz) & & & (kHz) & (km s$^{-1}$) \\
  \hline\noalign{\smallskip}
Ortho-{\rm{H$_2$O}} 1$_{10}$ -- 1$_{01}$ & 556.93599 & SWAS & 3.3$^{\prime }$ $\times $ 4.5$^{\prime }$$^a$ & $1.0\times 10^{3}$ & 0.55 \\
CS (2--1) & 97.98095 & SEST & 42$^{\prime \prime}$ & 43 & 0.13 \\
CS (5--4) & 244.93556 & SEST & 17$^{\prime \prime}$ & 43$^b$ & 0.052$^b$ \\
{\rm{N$_2$H$^+$}} (1--0) & 93.17304 & SEST & 44$^{\prime \prime}$ & 43 & 0.14 \\
  \hline\noalign{\smallskip}
\multicolumn{6}{l}{Notes: $^a$ \citealt{Melnick+etal+2000};} 
{$^b$ For G267.9--1.1, $\Delta v$ of{ the} CS (5--4) line is 0.060\
km s$^{-1}$, and $\Delta \nu $ is 49~kHz.}
\end{tabular}
\end{table*}

\subsubsection{SWAS observation}

The observations {of the} ortho-{\rm{H$_2$O}} 1$_{10}$ -- 1$_{01}$ line
(556.936~GHz) were performed with SWAS from 1999 January 20 to
2001 May 3 (G267.9--1.1), 1998 December 20 to 2003 June 5
(G268.4--0.9), 1999 September 15 to 2002 February 25 (G333.1--0.4)
and 2001 September 22 to 2004 July 21 (G336.5--1.5). The data were
obtained from the SWAS spectrum service in{ the} NASA/IPAC infrared
science archive\footnote{\it
http://irsa.ipac.caltech.edu/applications/SWAS/SWAS/list.html}.

The ortho-{\rm{H$_2$O}} 557~GHz 1$_{10}$ -- 1$_{01}$ line data
acquired by SWAS {were} converted into FITS format with a uniform 190
$\times $190~arcsec$^2$ pixel size after the spectra in every single
beam (which are {also }in the same sampling cell) were averaged and
then the baselines {were} substracted.
The Gildas software package\footnote{\it
http://www.iram.fr/IRAMFR/GILDAS/} was used for averaging and
baseline subtraction. The baselines of spectra {were} acceptable
and a 1$^{\rm st}$ or 2$^{\rm nd}$ order polynomial was used for
baseline fitting. The typical {root mean squares (}RMSs{)} of the
ortho-{\rm{H$_2$O}} 557~GHz 1$_{10}$ -- 1$_{01}$ spectra are
0.017~K for G267.9--1.1, 0.013~K for G268.4--0.9, 0.014~K for
G333.1--0.4, and 0.03~K for G336.5--1.5. The different RMSs are
mainly due to different integrat{ion} times. When we calculated
the integrated intensities of{ the} ortho-{\rm{H$_2$O}} 557~GHz
1$_{10}$ -- 1$_{01}$ line, the antenna temperatures were corrected
with a main beam efficiency of 0.9.

For G268.4--0.9 and G333.1--0.4, the antenna temperatures below
zero are due to the high noises and the subtraction of{ the} baseline.
The double-peaked spectra of ortho-{\rm{H$_2$O}} 1$_{10}$ -
1$_{01}$ lines of G268.4--0.9 indicated strong self-absorption
(\citealt{Ashby+etal+2000}). We performed{ a} Gaussian fitting for
both non-absorbed emission peak{s} and the absorption peak{s} and {obtained}
the integrated intensity of the emission of the averaged and
baseline subtracted spectrum. The spectrum of G333.1--0.4 shows a
pronounced inverse P{ }Cygni profile. We took into account both the
water components corresponding to emission and absorption
features.

\subsubsection{SEST observation}
The observations of{ the} CS $J = 2 \to  1$ line (97.98095~GHz), CS
$J = 5 \to  4$ line (244.93556~GHz) and {\rm{N$_2$H$^+$}} $J = 1\to
0${ line} (93.17632~GHz) were carried out with SEST. These four
clumps were mapped with CS $J = 2 \to  1$ (except for G333.1--0.4),
CS $J = 5 \to  4$ and {\rm{N$_2$H$^+$}} $J = 1\to  0$ in 2002 Mar
24--28. The main-beam efficiencies {were} 0.73 (CS $J = 2 \to  1$),
0.56 (CS $J = 5 \to  4$) (\citealt{Lapinov+etal+1998}) and 0.74
({\rm{N$_2$H$^+$}} $J = 1\to  0$) (\citealt{Mardones+etal+1997}),
respectively.

In the mappings, the spacing of the square scanning grids is $40''$,
but in the CS (5--4) mappings for G268.4--0.9, G333.1--0.4 and
G336.5--1.5, additional sampling made the square scanning grids
quincunxes.
 In each map, every pixel was
observed with the position switch mode separately. The reference
positions are selected approximate{ly} 1800$''$ away from the
centers of the maps (the coordinates in Columns~(2) and (3) 
of Table~\ref{complex}).

The Gildas software package was also used. For several spectra, the
baselines seem {to follow }a sine function but with changing periods
and amplitudes. In addition, for several spectra the line widths of
the real emission lines are similar to the periods of their sine
baselines, so we left them with their sinusoidal baselines. We check
all the spectra one by one for baseline fitting. These spectra are
near the edge of the mapping area and our calculation results are
affected{ little}.

After the baseline subtraction, the obtained spectra with RMSs less
than 1\,K (for CS $J = 5 \to 4$ in G333.1--0.4 and G336.5--1.5, the
sigma limits are 0.5~K and 0.84~K, respectively) were selected and
written in FITS format with a uniform $40 \times  40$~arcsec$^2$
pixel size.

The RMSs of the CS (2--1) spectra in the center of the images are
0.16~K for G267.9--1.1 (RA = 08:59:12.0, Dec = $-$47:29:04), 0.16~K
for G268.4--0.9 (RA = 09:01:54.3, Dec = $-$47:43:59) and   0.16~K for
G336.5--1.5 (RA = 16:39:58.9, Dec = $-$48:51:00). The RMSs of the CS
(5--4) spectra in the center of the images are 0.42~K for G267.9--1.1
(RA = 08:59:12.0, Dec = $-$47:29:04), 0.16~K for G268.4--0.9 (RA =
09:01:54.3, Dec = $-$47:43:59), 0.13~K for G333.1--0.4 (RA =
16:21:02.1, Dec = $-$50:35:15), and 0.09~K for G336.5--1.5 (RA =
16:39:58.9, Dec = $-$48:51:00).

We did not have CS (2--1) data for G333.1--0.4. The RMSs of the
{\rm{N$_2$H$^+$}} (1--0) spectra in the center of the images are
0.20~K for G267.9--1.1 (RA = 08:59:12.0, Dec = $-$47:29:04), 0.18~K
for G268.4--0.9 (RA = 09:01:54.3, Dec = $-$47:43:59), 0.19~K for
G333.1--0.4 (RA = 16:21:00.8, Dec = $-$50:34:55), and 0.17~K for
G336.5--1.5 (RA = 16:39:58.9, Dec = $-$48:51:00).

\subsubsection{CSO observation}
The 350~$\upmu$m dust continuum
 observations were performed with
 SHARC~II  
  on CSO during 2014
April 4 and 5. The data were taken when these four regions were close
to their maximum elevation (approximate{ly} 20 degree{s} at the CSO
site) and the $\tau_{225~ {\rm GHz}}$ was lower than 0.06. The box
scan mode was used for{ the} SHARC II observation. The beam size of
SHARC{ }II is 8$''$ and the grid spacing{ for the sampling} is 1.5
$\times$1.5~arcsec$^2$. For each scan, the total integration time is
14.71 minutes{ and} the corresponding RMS is 212~mJy~beam$^{-1}$.
Pointing and focusing calibration was done every 2 hours during the
observation. The data reduction tool CRUSH\footnote{\it
http://www.submm.caltech.edu/~sharc/crush/} was used for further data
reduction. The flux calibration was done {by} observing Mars. The RMS
for the final data are 0.49~Jy~beam$^{-1}$ for G267.9--1.1,
0.61~Jy~beam$^{-1}$ for G268.4--0.9, 0.79~Jy~beam$^{-1}$ for
G333.1--0.4 and 0.53~Jy~beam$^{-1}$ for G336.5--1.5. The weather is
the main reason for the variation of the noises in different maps.

\section{Results and Analysis}

\subsection{Spectra Map and Dust Map}

Figures~\ref{g267}, \ref{g268}, \ref{g333} and \ref{g336} are line
profile maps of G267.9--1.1, G268.4--0.9, G333.1--0.4 and
G333.6--1.5. In these line profile maps, all offsets are relative
to the corresponding coordinates ({s}ee Table~\ref{complex})
and the units are arcsec. Empty boxes are the
positions without sampling.

There is a ``hole'' with little CS (2--1 and 5--4) emission in the
center of the emission region of G267.9--1.1. {Moreover}, the
centroid velocities of the CS spectra in the east of the hole are
different {from} those of the CS spectra in the west of the hole. In
the south of the hole, the CS spectra all have two obvious peaks and
in the north of the hole, the spectra all have two peaks as well. We
overlapped the CS (2--1) integrated intensity map on the 350~$\upmu$m
dust continuum in Figure~\ref{hole}. We can see that the intensity
peaks are associated with the 350~$\upmu$m emission and in the
``hole'' the dust emission is much weaker than the surrounding areas.

The RMSs of CS (5{--}4) spectra of G333.1--0.4 var{y} a lot, which is
caused by different integration time{s}. The integration time (on
source time) changes from less than 0.8 minute{s} to more than 3
minutes. In G336.5--1.5's CS (5--4) spectra, there are similar
situations.

The 350~$\upmu$m dust continuums of these four clumps are shown in
Figure~\ref{G_350}. In the following sections, we estimated the
temperatures, masses, H$_2$ column densities and ortho-water
abundances of these four clumps. The areas for mass and ortho-water
abundance estimates are shown {in} white boxes with solid lines and
dashed lines, in the corresponding figures, respectively.

\begin{figure*}
\centering

\vs

\includegraphics[width=12cm]{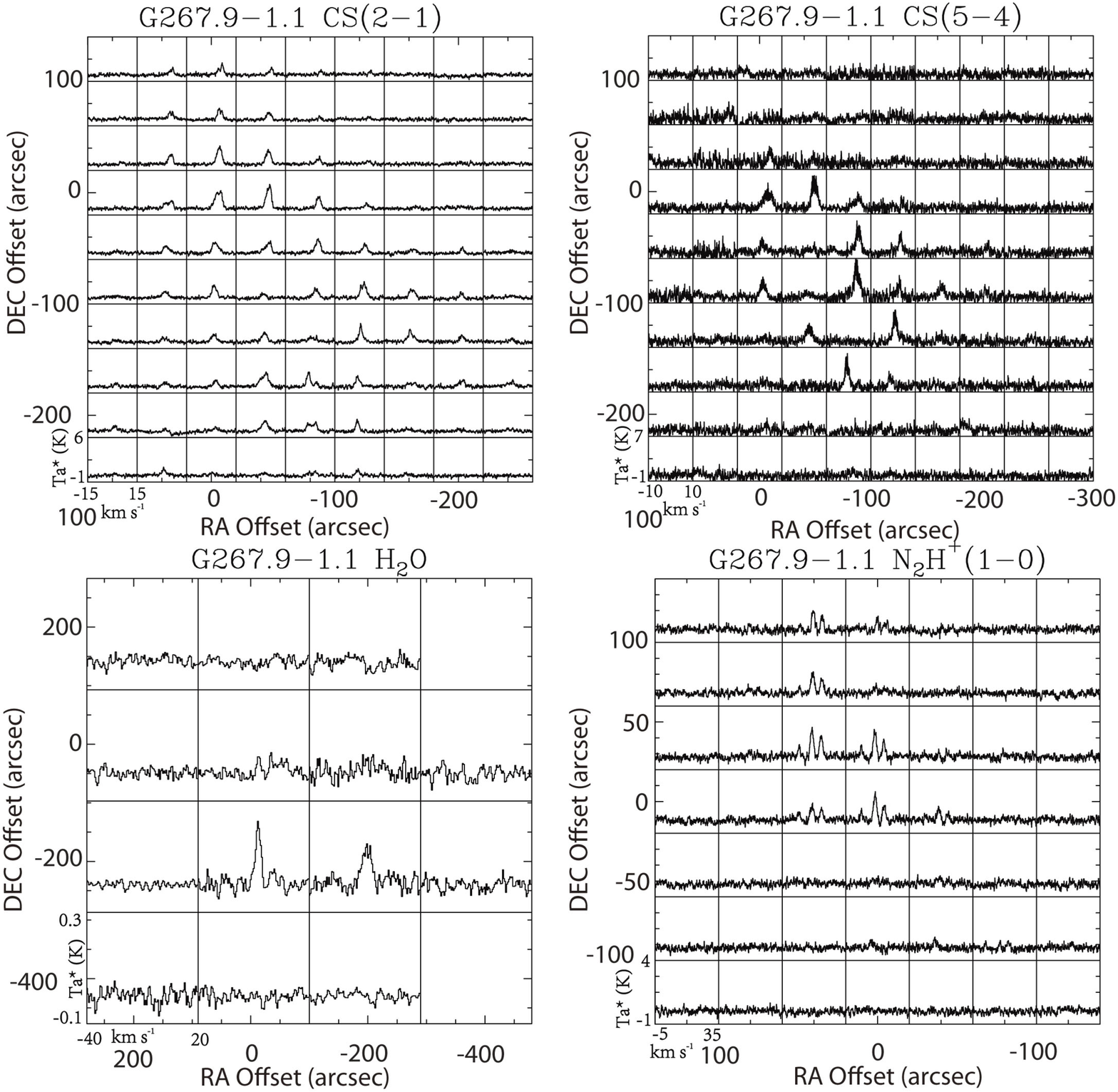}

 \caption{
\baselineskip 3.6mm
 {\it{Upper Left}}: G267.9--1.1 CS (2--1) line profile map.
It is obvious that at position{s} ($-40''$, $-80''$) and ($-80'',
-120''$) relative to the coordinates in Table~\ref{complex}, the
emissions at these two position{s} are much weaker than the
surrounding{s} and that may suggest ``holes'' in molecular gas
with significantly lower H$_2$ volume density. Similar and more
significant phenomena can be seen in the CS (5--4) line profile map.
 {\it{Upper Right}}: G267.9--1.1 CS (5{--}4{)} line profile map.
 {\it{Lower Left}}: G267.9--1.1 ortho-{\rm{H$_2$O}} 1$_{10}$--1$_{01}$ line profile map. The two empty boxes are positions without sampling.
 {\it{Lower Right}}: G267.9--1.1 {\rm{N$_2$H$^+$}} (1--0) line profile map.}
 \label{g267}
\end{figure*}
\begin{figure*}[tbp]
\centering

\includegraphics[width=12cm]{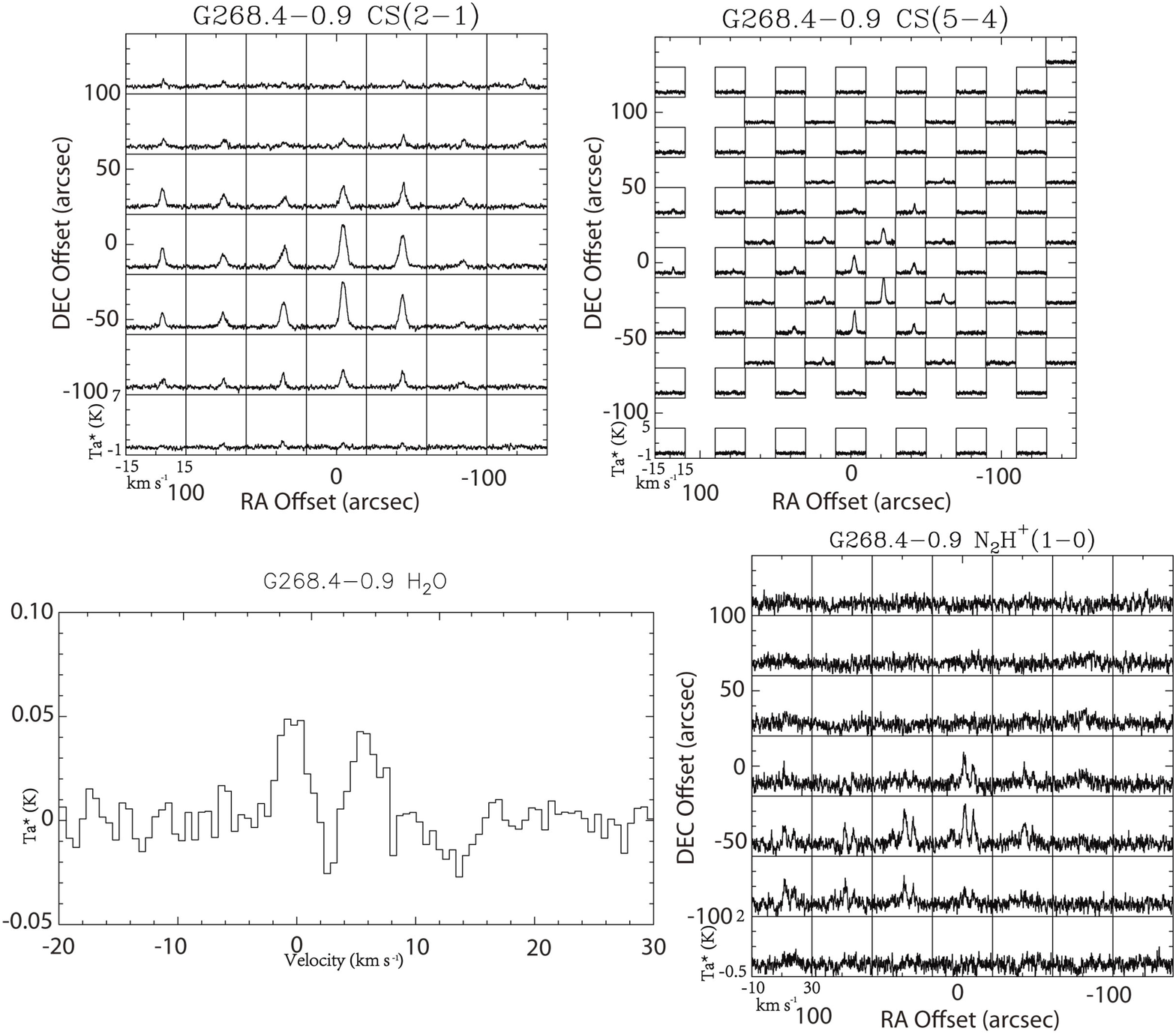}

 \caption{\baselineskip 3.6mm {\it{Upper Left}}: G268.4--0.9 CS (2--1) line profile map.
 {\it{Upper Right}}: G268.4--0.9 CS (5--4) line profile map. The empty boxes are positions without sampling.
 {\it{Lower Left}}: G268.4--0.9 ortho-{\rm{H$_2$O}} 1$_{10}$ -- 1$_{01}$ spectrum at RA 09:01:54.31, Dec --47:43:59.0.
 {\it{Lower Right}}: G268.4--0.9 {\rm{N$_2$H$^+$}} (1--0) line profile map.}
 \label{g268}

\end{figure*}
\begin{figure*}[tbp]
\centering

\includegraphics[width=12cm]{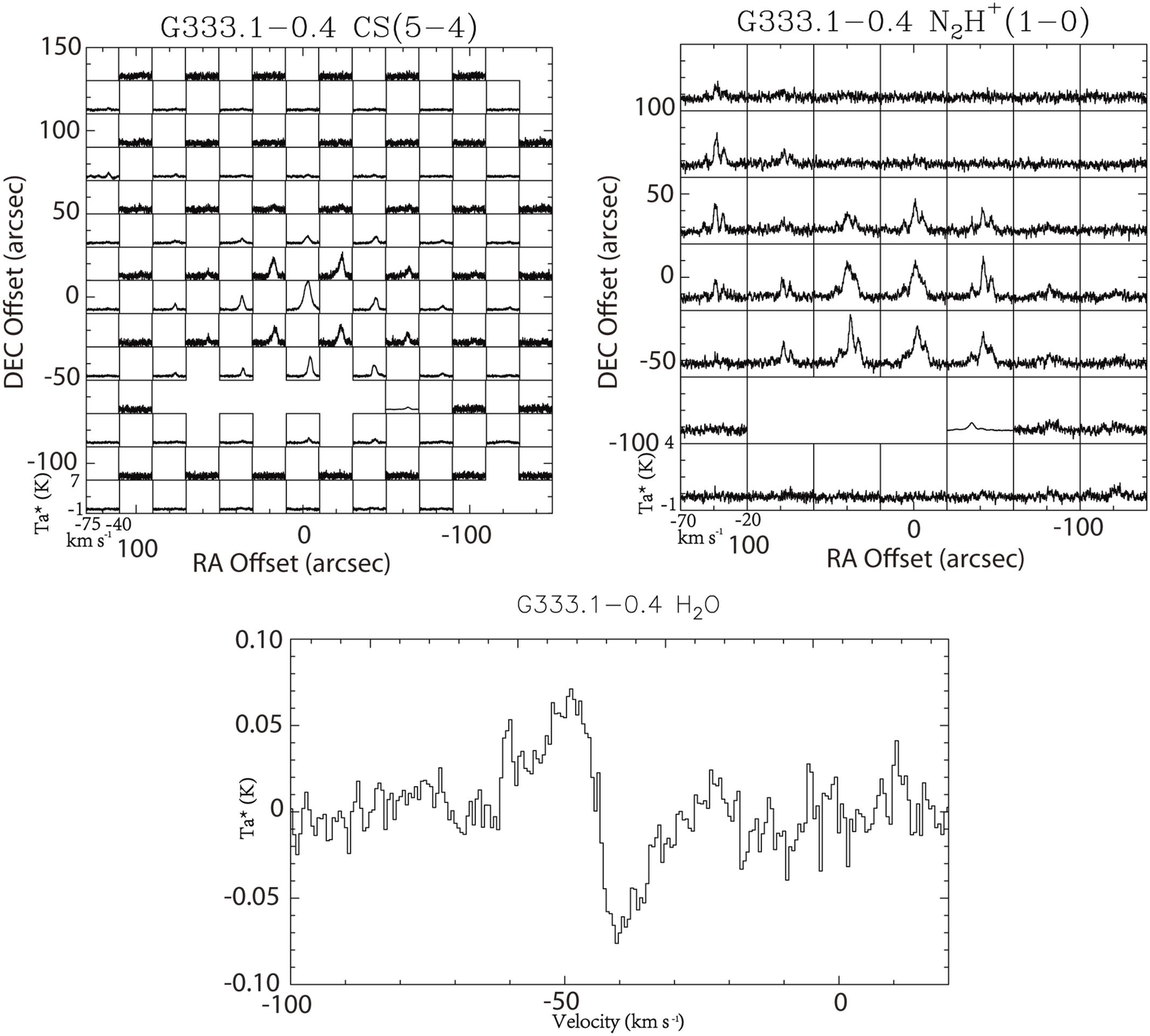}

 \caption{\baselineskip 3.6mm
 {\it{Upper Left}}: G333.1--0.4 CS (5--4) line profile map. The empty boxes are positions without sampling. The RMS changes a lot in different spectra.
 {\it{Upper Right}}: G333.1--0.4 {\rm{N$_2$H$^+$}} (1--0) line profile map. A spectrum with extremely low RMS is shown. The three empty boxes are positions without sampling.
 {\it{Lower}}: G333.1--0.4 ortho-{\rm{H$_2$O}} 1$_{10}$ -- 1$_{01}$ spectrum at RA 16:21:02.08, Dec --50:35:15.0.}
 \label{g333}
\end{figure*}
\begin{figure*}[tbp]
\centering
\includegraphics[width=12cm]{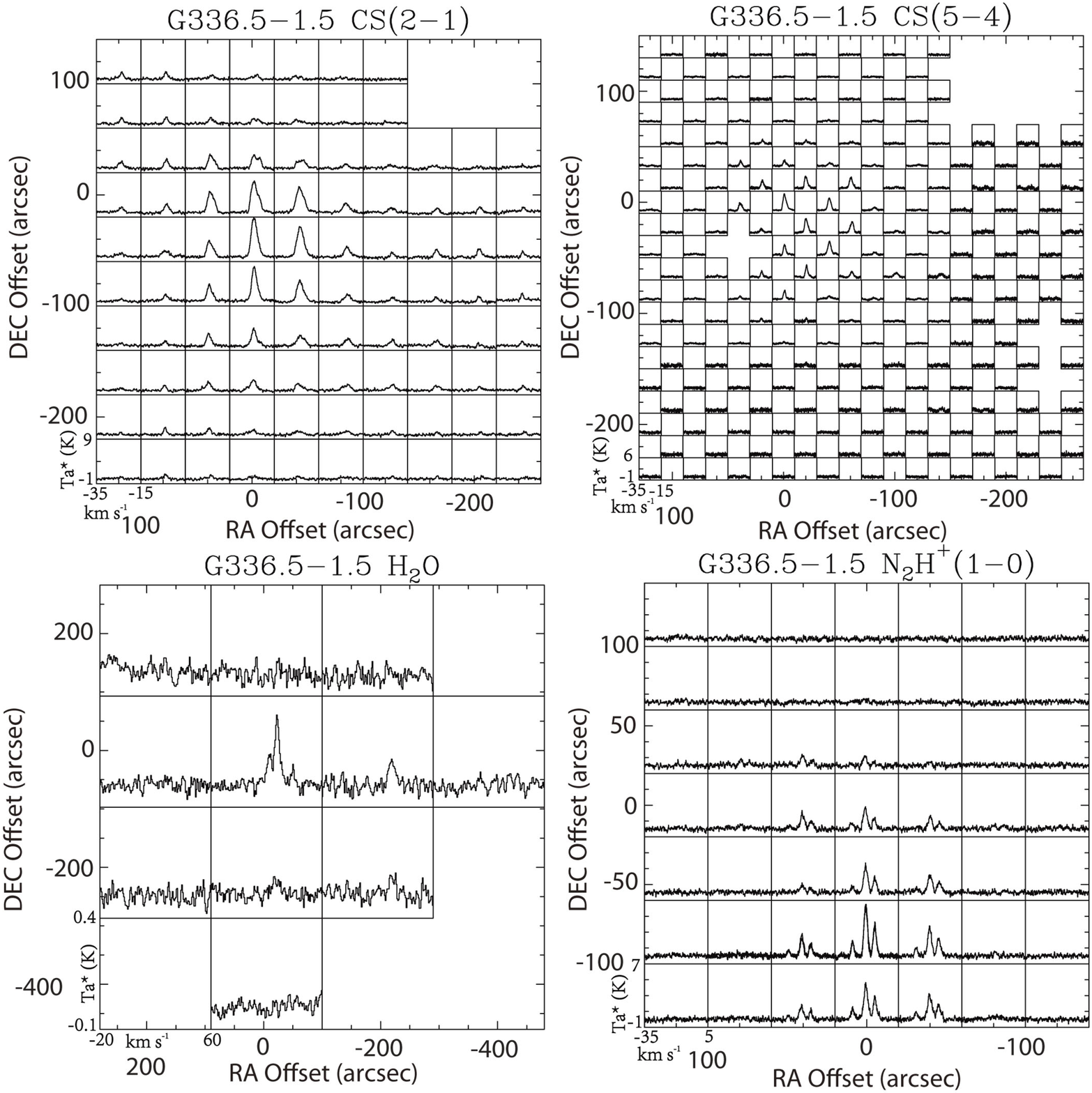}

 \caption{\baselineskip 3.6mm  {\it{Upper Left}}: G336.5--1.5 CS (2--1) line profile map.  The empty boxes at the upper right corner are positions without sampling.
 {\it{Upper Right}}: G336.5--1.5 CS (5--4) line profile map.  The empty boxes are positions without {an }observation.
 {\it{Lower Left}}: G336.5--1.5 ortho-{\rm{H$_2$O}} 1$_{10}$ -- 1$_{01}$ line profile map. The empty boxes are positions without sampling.
 {\it{Lower Right}}: G336.5--1.5 {\rm{N$_2$H$^+$}} (1--0) line profile map.}
 \label{g336}
\end{figure*}

\begin{figure*}
\centering

\includegraphics[width=7.1cm]{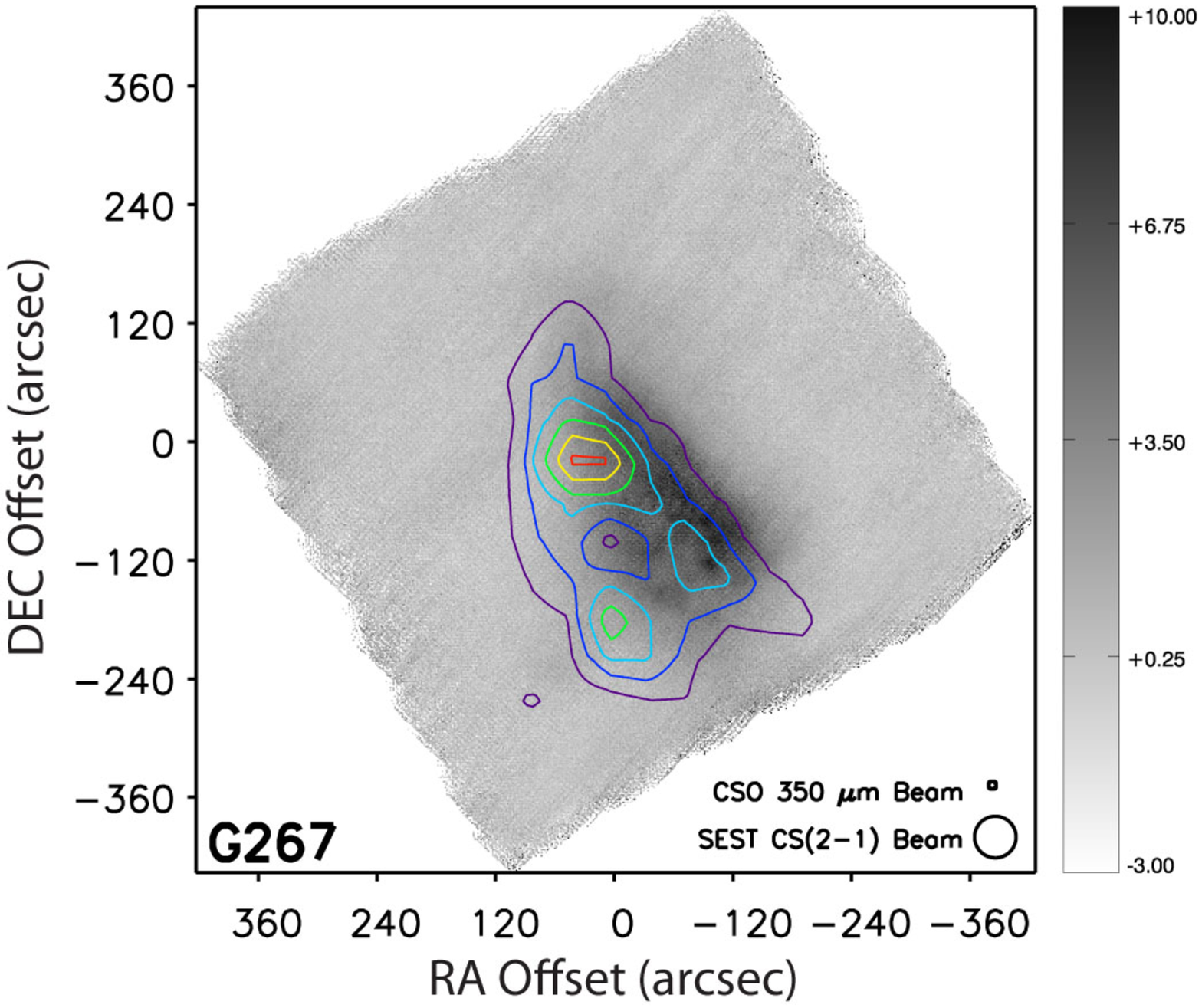}

 \caption{
\baselineskip 3.6mm
 The ``hole'' in G267.9--1.1. The background
grey scale map is the 350~$\upmu$m emission map of G267.9--1.1 and
the color contours show the CS (2--1) integrated intensity. The
purple, dark blue, light blue, green, yellow and red contours
represent 3, 5, 7, 9, 11 and 13~K~km~s$^{-1}$, respectively. The
flux values on the color bar are in unit{s} of Jy~beam$^{-1}$.}
 \label{hole}
\centering

\vs\vs

\includegraphics[width=12cm]{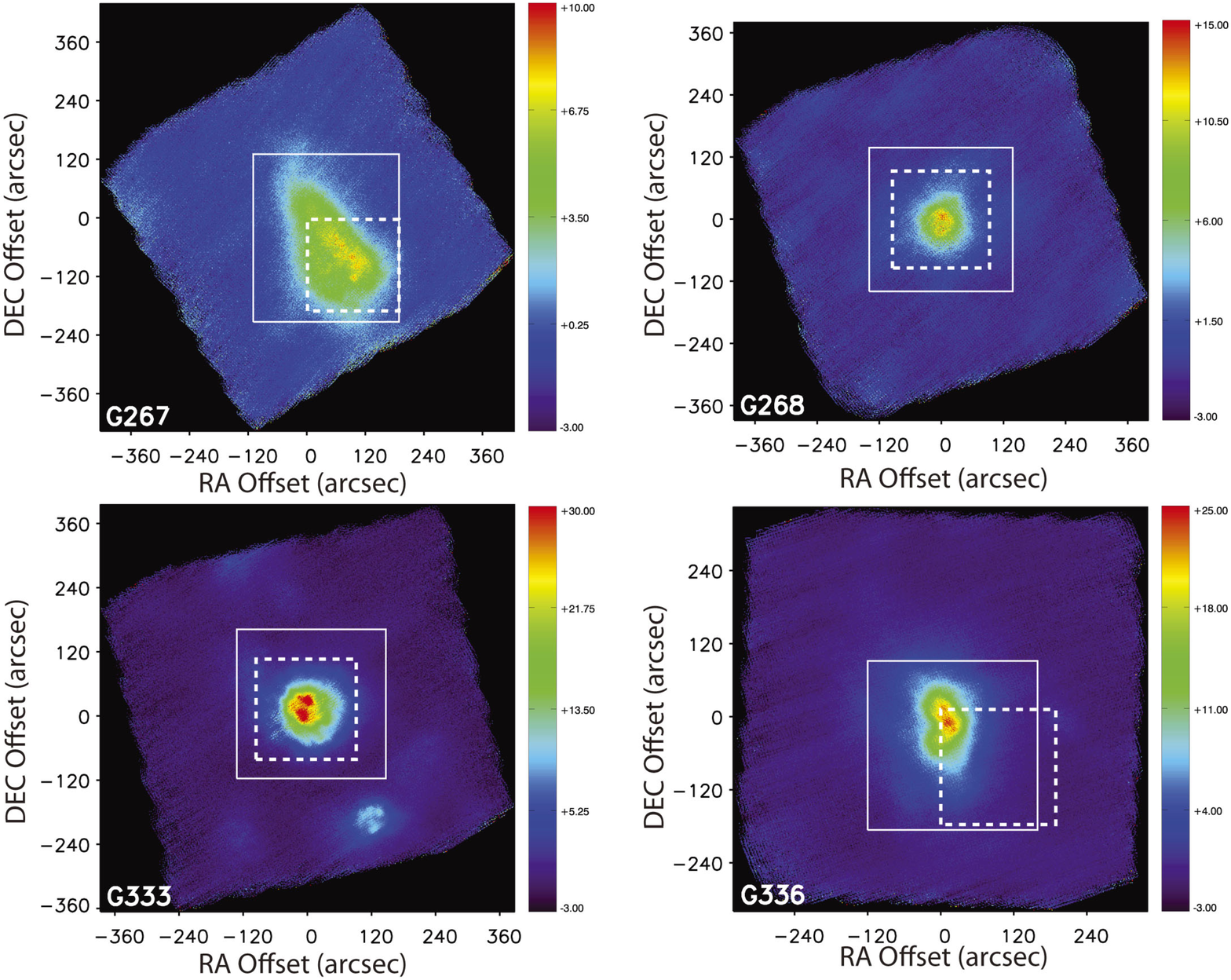}

 \caption{
\baselineskip 3.6mm
 The white boxes with solid lines and dashed lines represent the mass
 estimate area (also the CS (5--4)-traced area) and the ortho-{\rm{H$_2$O}}
abundance estimate area, respectively. The flux values on the
color bar are in unit{s} of Jy~beam$^{-1}$.
 {\it Upper Left}: G267.9--1.1 350~$\upmu $m map.
 {\it Upper Right}: G268.4--0.9 350~$\upmu $m map.
 {\it Lower Left}: G333.1--0.4 350~$\upmu $m map.
 {\it Lower Right}: G336.5--1.5 350~$\upmu $m map.}
 \label{G_350}
\end{figure*}

\subsection{CS Excitation Temperatures}

\subsubsection{The estimate of CS excitation temperatures}

For G267.9--1.1, G268.4--0.9 and G336.5--1.5 (we did not have CS
(2{--}1) line data for G333.1--0.4) the excitation temperatures of
{the }CS molecule were estimated based on{ the} CS (2--1) line and
CS (5--4) line. The estimated CS excitation temperatures were
subsequently adopted in the estimates of clump masses (and then
the H$_2$ column densities), ortho-water abundances and N$_2$H$^+$
abundances.

The estimate is based on {the }following assumptions, i.e., (1)
The
CS molecules are in LTE. 
(2) The cosmic microwave background radiation (CMB) can be
ignored. (3) CS (2--1) and CS (5--4) lines are optically thin.
Since G267.9--1.1, G268.40.9 and G336.5--1.5 all fill the main
beam, the filling factors in the estimate equal 1. According to
the population diagram method (in LTE)
(\citealt{Goldsmith+Langer+1999}), {f}or the upper levels $J = 2$
and $J = 5$ there are
%
%
\begin{equation}\label{eq1}
\ln \frac{N_{\rm{J=2}}^{\rm{thin}}}{g_{\rm{J=2}}}+\ln \frac{\tau _{2-1}}{1-e^{\tau _{2-1}}}=\ln N_{\rm{tot}}-\ln Z-\frac{E_{\rm{J=2}}}{kT_{\rm{ex}}},
\end{equation}
and
\begin{equation}\label{eq2}
\ln \frac{N_{\rm{J=5}}^{\rm{thin}}}{g_{\rm{J=5}}}+\ln \frac{\tau _{5-4}}{1-e^{\tau _{5-4}}}=\ln N_{\rm{tot}}-\ln Z-\frac{E_{\rm{J=5}}}{kT_{\rm{ex}}},
\end{equation}
respectively. \textit{$N$}$_{\rm{J=2}}^{\rm{thin}}$ and
\textit{$N$}$_{\rm{J=5}}^{\rm{thin}}$ are the column densities at $J
= 2$ and $J = 5$ in {the }optically thin situation{ respectively}.
$g$$_{\rm{J=2}}$ and $g$$_{\rm{J=5}}$ are the statistical weights of
level $J = 2$ and level $J= 5${ respectively},{ and} $\tau$$_{5-4}$ and $\tau$$_{2-1}$
are the corresponding optical depths. $N$$_{\rm{tot}}$ is the total
column density of {the }CS molecule and $Z$ is the partition
function. $T$$_{\rm{ex}}$ is the excitation temperature of these two
transitions. Since we assumed that{ the} CS (2--1) line and CS (5--4)
line are optically thin, we considered
\begin{equation}\label{eq3}
\ln \frac{\tau _{2-1}}{1-e^{\tau _{2-1}}}=0
\end{equation}
and
\begin{equation}\label{eq4}
\ln \frac{\tau _{5-4}}{1-e^{\tau _{5-4}}}=0.
\end{equation}
{F}rom Equations~(\ref{eq1}), (\ref{eq2}), (\ref{eq3})
and (\ref{eq4}), we obtained
\begin{equation}\label{eq5}
T_{\rm{ex}}=\frac{E_{\rm{J=5}}-E_{\rm{J=2}}}{k\ln
 \left ( \frac{g_{\rm{J=5}}N_{\rm{J=2}}^{\rm{thin}}}{g_{\rm{J=2}}N_{\rm{J=5}}^{\rm{thin}}} \right )},
\end{equation}
while
\begin{equation}\label{eq6}
E_{J}=hB_{\rm{e}}J(J+1).
\end{equation}
$J$ is the rotational quantum number{ and} $B$$_{\rm{e}}$ is the
rotational constant of{ the} CS molecule at vibrational energy
 level $v$
= 0 in Hz ($2.458437 \times 10^{10}$~Hz, \citealt{Kewley+etal+1963}).
The statistical weights of{ the} $J = 2$ and $J = 5$ level,
$g$$_{\rm{J=2}}$ and $g$$_{\rm{J=5}}$, equal 5 and 11 respectively.
Thus, we can write Equation~(\ref{eq5}) as
\begin{equation}\label{eq7}
T_{\rm{ex}}=\frac{24hB_{\rm{e}}}{ k\ln \left ( \frac{11N_{\rm{J=2}}^{\rm{thin}}}{5N_{\rm{J=5}}^{\rm{thin}}} \right )}.
\end{equation}

Now we focus on the CS column densities of the upper levels ($J = 2$
and $J = 5$). Based on \cite{Rohlfs+Wilson+1996}, when the molecular
line is optically thin, the column densities of {the }upper
level ($J = 2$ or $J = 5$) {are}
\begin{equation}\label{eq8}
N_{\rm{u}}^{\rm{thin}}=\frac{8\pi \nu ^{3}}{c^{3}A_{\rm{ul}}\left ( e^{\frac{h\nu }{kT_{\rm{ex}}}}-1 \right )}\int \tau _{\nu }dv,
\end{equation}
{w}here $\nu $ is the frequency of{ the} CS (2--1) or CS (5--4)
line, $T$$_{\rm{ex}}$ is the excitation temperature,
$A$$_{\rm{ul}}$ is the corresponding Einstein $A$-coefficient and
$\tau $$_{\nu }$ is the optical depth. In {an }isothermal medium,
the relationship between the brightness temperature $T_{\rm B}$,
the excitation temperature $T$$_{\rm{ex}}$ and
the cosmic background temperature $T$$_{\rm{background}}$ can be
describe{d} as
\begin{equation}\label{eq9}
T_{\rm B}
=T_{{\rm{background}}}e^{-\tau _{\rm{\nu
}}}+T_{\rm{ex}}\left ( 1-e^{\tau _{\rm{\nu }}} \right ).
\end{equation}
When the CS lines {are} optically thin,
\begin{equation}\label{eq10}
1-e^{-\tau _{\rm{\nu }}}\approx \tau _{\nu }{\,,}
\end{equation}
and if the background radiation can be ignored, then
\begin{equation}\label{eq11}
T_{\rm B}
\approx T_{\rm{ex}}\tau _{\rm{\nu }}.
\end{equation}
Substituting Equation~(\ref{eq11}) into Equation~(\ref{eq8}), we
{obtained}
\begin{eqnarray}\label{eq12}
N_{\rm{u}}^{\rm{thin}}&=&\frac{8\pi \nu
^{3}}{c^{3}A_{\rm{ul}}T_{\rm{ex}} \left ( e^{\frac{h\nu
}{kT_{\rm{ex}}}}-1 \right )} \int T_{\rm{ex}}\tau _{\nu }dv
\nonumber \\ &=&\frac{8\pi \nu
^{3}}{c^{3}A_{\rm{ul}}T_{\rm{ex}}\left ( e^{\frac{h\nu
}{kT_{\rm{ex}}}}-1 \right )}\int T_{\rm B}
dv,
\end{eqnarray}
Since the four clumps in our study are all extended sources, the
antenna temperature $T$$_{\rm{a}}$ $\simeq $ $T_{\rm
B}
$, {and} the column densities of upper levels in {the }optically
thin {case} {are}
\begin{equation}\label{eq13}
N_{\rm{u}}^{\rm{thin}}=\frac{8\pi \nu ^{3}}{c^{3}A_{\rm{ul}}T_{\rm{ex}}\left ( e^{\frac{h\nu }{kT_{\rm{ex}}}}-1 \right )}\int T_{\rm{a}}dv.
\end{equation}
In Equation~(\ref{eq13}), $N$$_{\rm{u}}^{\rm{thin}}$ is dependent on $T$$_{\rm{ex}}$.

If we adopt the Rayleigh-Jeans approximation
\begin{equation}\label{eq14}
h\nu \ll kT_{\rm{ex}}
\end{equation}
in Equation~(\ref{eq13}), then this equation will be reduced to
\begin{equation}\label{eq15}
N_{\rm{u}}^{\rm{thin}\ast }=\frac{8\pi k\nu ^{2}}{hc^{3}A_{\rm{ul}}}\int T_{\rm{a}}dv.
\end{equation}
This expression is the same as that from \cite{Goldsmith+Langer+1999}, and it does not depend on the excitation temperature $T$$_{\rm{ex}}$.

However, we notice that if we estimate the column densities of upper
levels through Equation~(\ref{eq15}), then significant deviation
will arise due to the high frequency of {the }CS (5--4) line.
The deviation in upper level column densities will lead to
significant deviation of the subsequently derived $T$$_{\rm{ex}}$.

The excitation-temperature-dependent $N$$_{\rm{u}}^{\rm{thin}}$ was derived by correcting the $N$$_{\rm{u}}^{\rm{thin}\ast }$ with line frequency $\nu $ and excitation temperature $T$$_{\rm{ex}}$
\begin{equation}\label{eq16}
N_{\rm{u}}^{\rm{thin}}=\frac{h\nu }{kT_{\rm{ex}}\left ( e^{\frac{h\nu }{kT_{\rm{ex}}}}-1 \right )}\cdot N_{\rm{u}}^{\rm{thin}\ast }.
\end{equation}

For each sampling position, we need $N$$_{\rm{u}}^{\rm{thin}}$ at{
the} $J = 2$ and $J = 5$ level to calculate the excitation
temperature of CS, while the excitation temperature of {the }CS
molecule is required when we derived $N$$_{\rm{u}}^{\rm{thin}}$
(Eq.~(\ref{eq16})). As a start, we calculated
$N$$_{\rm{u}}^{\rm{thin}\ast }$ for{ the} $J = 2$ and $J = 5$ level
(Eq.~(\ref{eq15})) and then we derived an excitation temperature
(Eq.~(\ref{eq7})) from $N$$_{\rm{u}}^{\rm{thin}\ast }$ ($J = 2$ and
$J =5$ level). {In the n}ext step, we {obtained} the first
$N$$_{\rm{u}}^{\rm{thin}}$ through Equation~(\ref{eq16}), with which
we subsequently calculated an excitation temperature again
(Eq.~(\ref{eq7})). Since $N$$_{\rm{u}}^{\rm{thin}}$ is dependent on
$T$$_{\rm{ex}}$, and $T$$_{\rm{ex}}$ is derived from
$N$$_{\rm{u}}^{\rm{thin}}$s ($J$ = 2 and $J$ = 5 level), we
performed iterative calculations to correct $T$$_{\rm{ex}}$
gradually, i.e., we ran the {iteration cycles} of ``$N_{\rm u}^{\rm
thin}$ -- $T_{\rm ex}${.}'' We kept comparing the latest
$T$$_{\rm{ex}}$ calculated from the latest
$N$$_{\rm{u}}^{\rm{thin}}$s at{ the} $J$ = 2 and $J$ = 5 level with
the previous{ly} calculated $T$$_{\rm{ex}}$. The iteration{s} would
not end until the absolute value of the difference {between} these
two $T$$_{\rm{ex}}$ values {was} less than 0.05~K.

The calculation of CS excitation temperatures {was} only applied {to}
the positions where both the CS (2--1) and CS (5--4) line intensities
{were} higher than 3$\sigma $. At these positions, gas along the
{line of sight} {would} be considered CS (5--4)-traced dense gas and
the areas correspond{ing} to all these positions are treated as ``CS
(5--4)-traced areas{,}'' in which the CS molecules can be
well-excited through collisions (at least for the CS $J$ = 1 $\to $ 2
excitation, since the critical density of{ the} CS (5--4) line is far
greater than that of the CS (2--1) line) and the calculated CS
excitation temperatures approximately equal the kinetic temperatures
of the gas{;} the kinetic temperature of the gas is an essential
parameter in subsequent estimates. Considering the sampling spacing
({comparing} it with the beam size), no interpolation was applied to
positions where the quality of signal is poor (low signal-to-noise
ratio) and the corresponding CS column densities at $J$ = 2 and $J$ =
5 as well as the $T_{\rm{ex}}$s were denoted as zero at these
positions. The average CS excitation temperatures of these four
clumps (corresponding to the areas within the white boxes (solid
lines) in Fig.~\ref{G_350}) are listed in Column (5) of
Table~\ref{complex}. In those areas all calculated CS excitation
temperatures are non-zero. We {subsequently }estimated the masses of
the clumps within these white boxes (solid lines) in Section~3.3.

We also derived {a} ``characteristic'' CS excitation temperature,
\textit{$T$}$_{\rm{ex-clump}}$, for the CS (5--4)-traced area in each
clump (G267.9--1.1, G268.4--0.9 and G336.5--1.5). For each clump, we
averaged the integrated intensity of{ the} CS (2--1) line (and CS
(5--4) line, as well) at every sample position in the CS
(5--4)-traced area ({t}he white boxes (solid lines) in
Fig.~\ref{G_350}), then with these average integrated intensities we
derived the ``characteristic'' excitation temperatures
\textit{$T$}$_{\rm{ex-clump}}$ through exactly the same assumptions
and processes as we previously described.

Figure~\ref{T_ex} shows the population diagrams calculated for the
\textit{$T$}$_{\rm{ex-clump}}$s of G267.9--1.1, G268.4--0.9 and
G336.5--1.5. The population diagrams were deduced from corresponding
$N$$_{\rm{u}}^{\rm{thin}\ast }$s (at $J$ = 2 and $J$ = 5, in dashed
lines) and the last $N$$_{\rm{u}}^{\rm{thin}}$s (at $J$ = 2 and $J$ =
5, in solid lines) we {retrieved} in the iteration. We can see that
the correction of the upper level column densities ($J$ = 2 and $J$ =
5) for the Rayleigh-Jeans approximation does make{ a} difference in
the slopes of these plots.  We derived the final values for the
\textit{$T$}$_{\rm{ex-clump}}$s, {i.e.}, those deduced from the
slopes of the plots in solid lines in Figure~\ref{T_ex}, were
listed in Table~\ref{NLTE-new}, 
 together with
\textit{$N$}$_{\rm{CS-clump}}$, the average CS column density in
the CS (5--4)-traced area of each clump. The average CS column
density in the CS (5--4)-traced area was deduced from the final
results of the iteration for \textit{$T$}$_{\rm{ex-clump}}$.

\begin{figure*} 
\centering

\includegraphics[width=8cm]{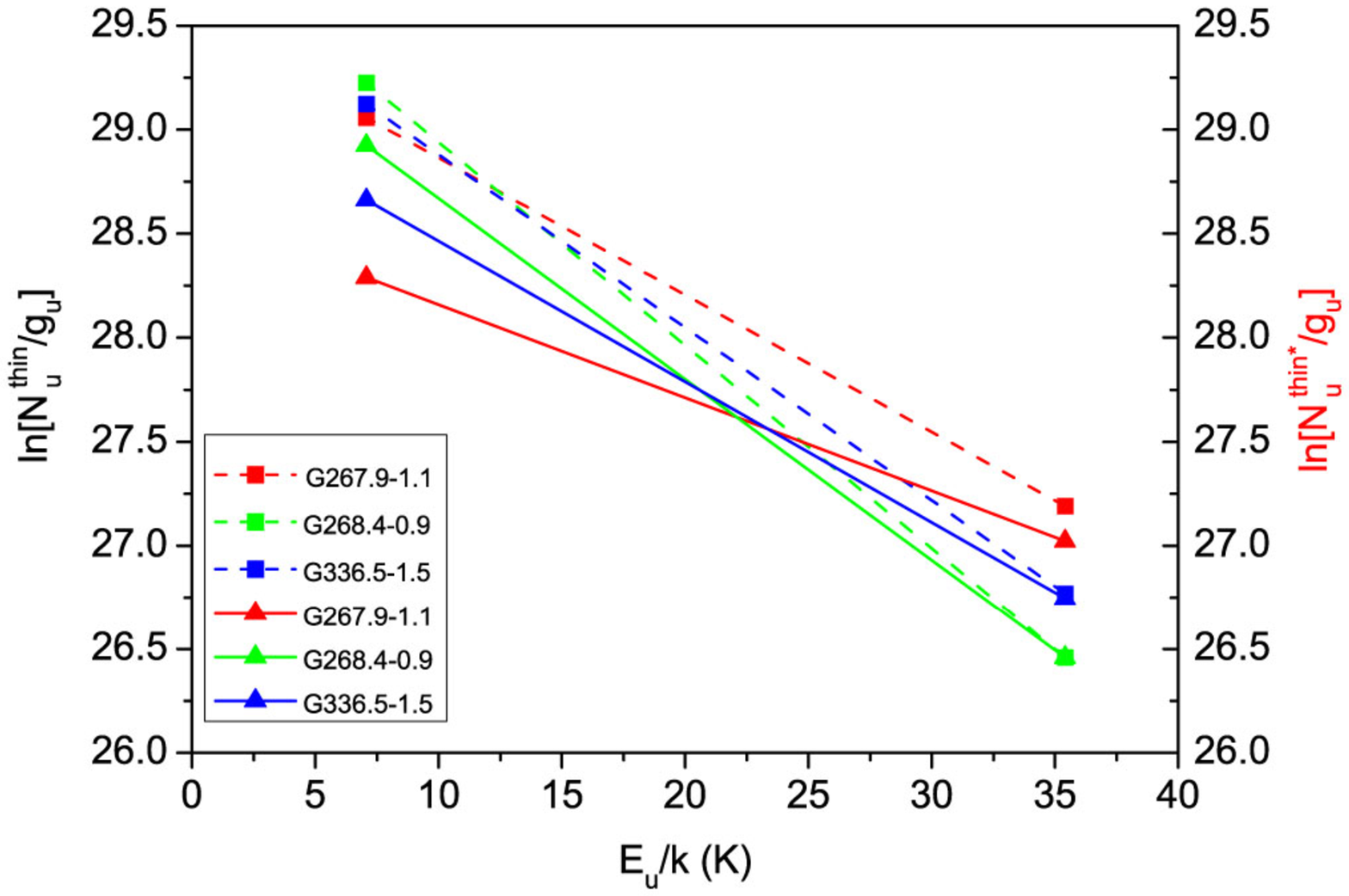}

 \caption{
\baselineskip 3.6mm Population diagrams for the ``characteristic''
CS excitation temperatures, $T_{\rm{ex-clump}}$s, of the CS
(5--4)-traced areas in G267.9--1.1, G268.4--0.9 and G336.5--1.5. The
plots in dashed lines with solid square symbols and in solid lines
with solid triangle symbols were deduced from
$N$$_{\rm{u}}^{\rm{thin}\ast}$s (Eq.~(\ref{eq15})) and the last
$N$$_{\rm{u}}^{\rm{thin}}$s
({Eq.}~{(}\ref{eq16})) {retrieved} in the iterations,
respectively. The red, green and blue plots represent G267.9--1.1,
G268.4--0.9 and G336.5--1.5, respectively.}
 \label{T_ex}
\end{figure*}
\begin{figure*}

\vs \centering

\includegraphics[width=9.55cm]{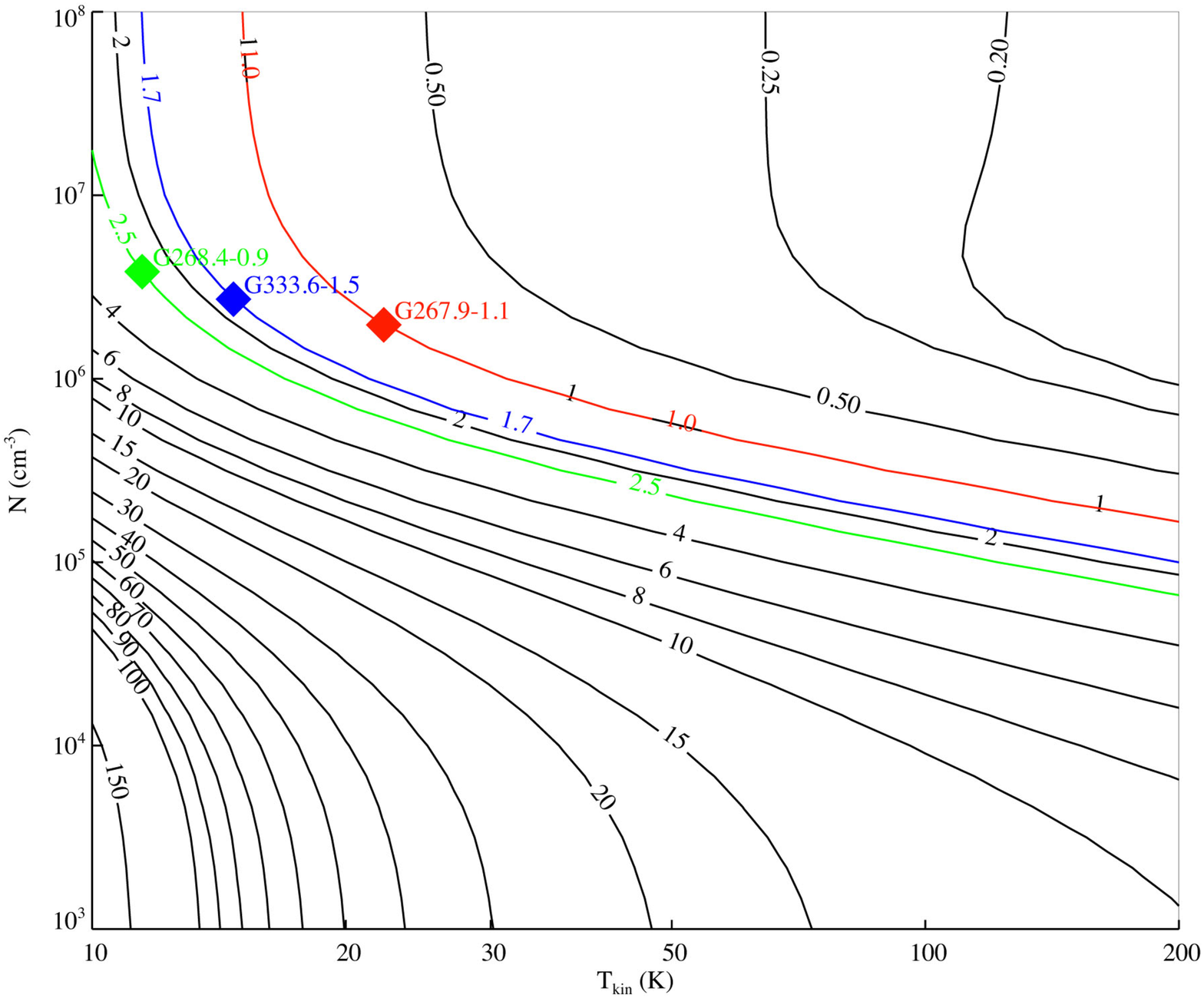}

\caption{ \baselineskip 3.6mm Line intensity ratio map on the kinetic
temperature-volume density plane with CS column densities {of}
10$^{13}$~cm$^{-2}$. The red, green and blue diamond{s} represent
G267.9--1.1, G268.4--0.9 and G336.5--1.5 respectively.}
 \label{NLTE_1}
\end{figure*}

\subsubsection{The {n}on-LTE analysis for CS (2--1) and CS (5--4) lines}

Since {it is possible for the CS lines} to be optically thick
(especially for the CS (2--1) line) in the actual situation, we also
performed non-LTE analysis for{ the} CS (2--1) and CS (5--4) lines
with RADEX (\citealt{vanderTak+etal+2007}) in the CS (5--4)-traced
areas (the white boxes with solid lines in Fig.~\ref{G_350}). To
perform this analysis,
 for each clump we added up the integrated intensities of{ the}
CS (2--1) line and CS (5--4) line at every sample position in the CS
(5--4)-traced area ({t}he white boxes (solid lines) in
Fig.~\ref{G_350}) respectively and {found} the corresponding average
values. Based on the average CS (2--1) and CS (5--4) integrated
intensities we {derived} the
 line intensity ratio
{\textit{$R$}$_{\rm{(2-1)/(5-4)}}$ (CS (2--1)/CS (5--4))} in the CS
(5--4)-traced area for each clump. We then performed the non-LTE
analysis with RADEX in a kinetic temperature range from 10\,K to
200~K and an H$_2$ density range from 10$^3$ to 10$^8$~cm$^{-3}$.
According to the \textit{$N$}$_{\rm{CS-clump}}$ we had estimated for
each clump (see Section 3.2.1), we performed the analysis with CS
column densities of 10$^{13}$~cm$^{-2}$ and 10$^{14}$~cm$^{-2}$ and
got the line intensity ratio maps on the kinetic temperature
(\textit{$T$}$_{\rm{kin}}$)-density plane. To estimate the {probable}
H$_2$ density within the CS (5--4)-traced area for each clump, we
need the corresponding kinetic temperature {in addition to} the
\textit{$R$}$_{\rm{(2-1)/(5-4)}}$. When we took their
\textit{$T$}$_{\rm{ex-clump}}$s as the corresponding kinetic
temperatures in the CS (5--4)-traced areas (and this is in accordance
with the LTE assumption which we would adopt in the following
estimates), the clumps are marked on the maps as
Figures~\ref{NLTE_1} and 
\ref{NLTE_2} show. The corresponding average H$_2$ densities of
the clumps within the CS (5--4)-traced areas,
\textit{$n$}$_{\rm{av}}$s, are listed in Table~\ref{NLTE-new}
together with other parameters. The non-LTE analysis suggests that
all these clumps have quite high \textit{$n$}$_{\rm{av}}$s in the
CS (5--4)-traced areas. The non-LTE analysis results offer
references for assuming densities in the subsequent estimates. We
{did not} perform non-LTE analysis for G333.1--0.4 since we only
{obtained} CS (5--4) data.

\begin{table*}
\centering

\begin{minipage}{80mm}

\caption{Non-LTE Analysis Parameters And Results} \label{NLTE-new}
\end{minipage}

\fns\tabcolsep 2mm
 \begin{tabular}{lccccc}
  \hline\noalign{\smallskip}
Clump      &  \textit{$R$}$_{\rm{(2-1)/(5-4)}}$ & \textit{$T$}$_{\rm{ex-clump}}$ (K) & \textit{$N$}$_{\rm{CS-clump}}$ (cm$^{-2}$) & \textit{$n$}$_{\rm{av}}$ (cm$^{-3}$)& \textit{$n$}$_{\rm{av}}$(cm$^{-3}$) \\
 & & & & \textit{$N$}$_{\rm{CS-clump}}$ = 10$^{13}$~cm$^{-2}$ & \textit{$N$}$_{\rm{CS-clump}}$ = 10$^{14}$~cm$^{-2}$ \\
  \hline\noalign{\smallskip}
G267.9--1.1 & 1.0 & 22.4 & $5.1\times 10^{13}$ & $2.0\times 10^{6}$ & $2.9\times 10^{6}$ \\
G268.4--0.9 & 2.5 & 11.5 & $6.8\times 10^{13}$ & $3.8\times 10^{6}$ & $1.2\times 10^{6}$ \\
G333.1--0.4$^{\rm a}$ & -- & -- & -- & -- & --            \\
G336.5--1.5 & 1.7 & 14.8 & $5.9\times 10^{13}$ & $2.7\times 10^{6}$ & $1.7\times 10^{6}$ \\
  \hline\noalign{\smallskip}
\multicolumn{6}{l}{Notes: $^{\rm a}$  We did not perform non-LTE analysis for clump G333.1-0.4 since there {were} only CS{ }(5{--}4) data.} \\
\end{tabular}
\end{table*}

\begin{figure*}[tbp]
\centering

\includegraphics[width=9.5cm]{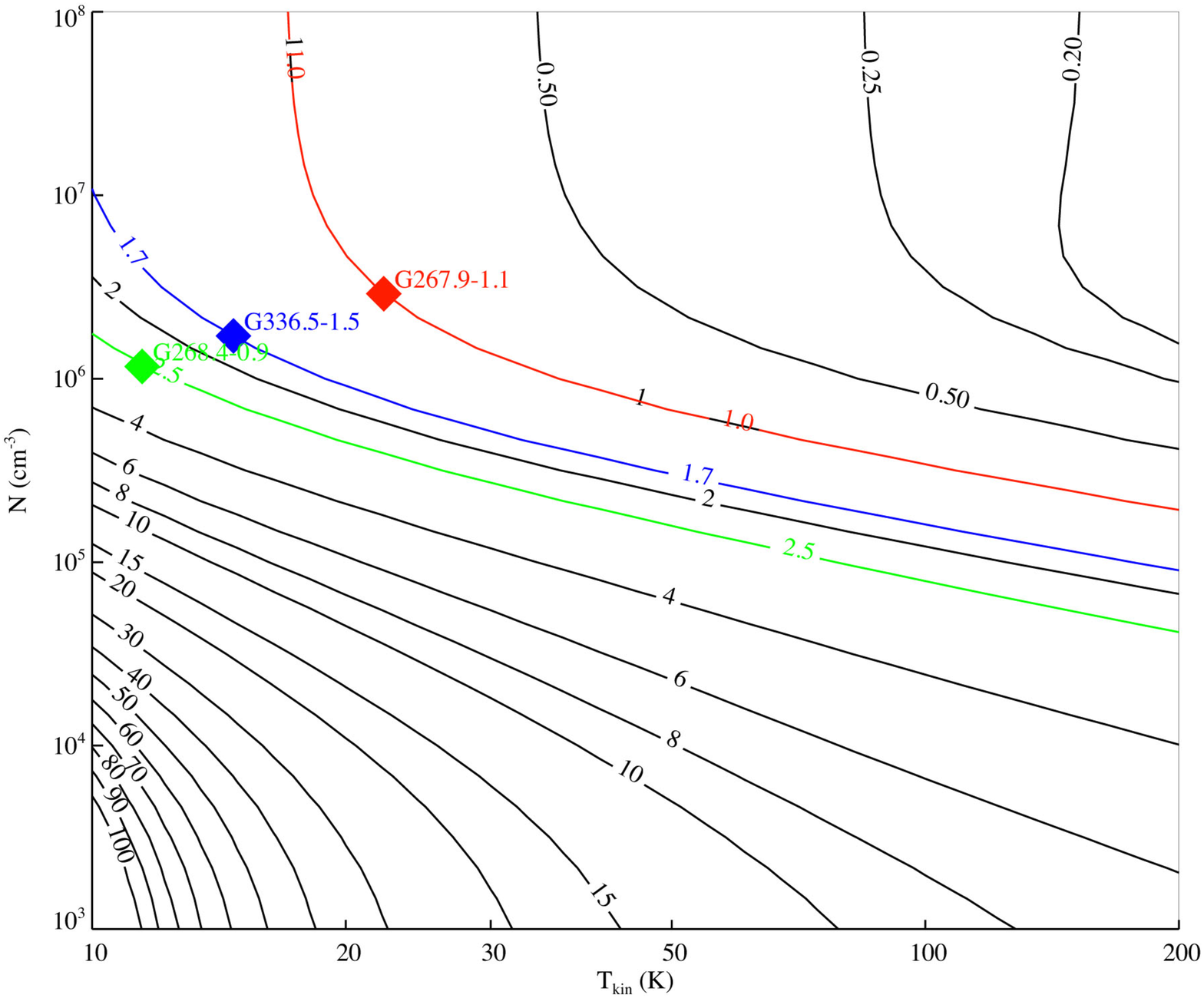}

\caption{ \baselineskip 3.6mm Line intensity ratio map on the
kinetic temperature-volume density plane with CS column densities
{of} 10$^{14}$~cm$^{-2}$. The red, green and blue diamond{s}
represent G267.9--1.1, G268.4--0.9 and G336.5--1.5 respectively. }
 \label{NLTE_2}
\end{figure*}

\subsection{Clump Masses and H$_2$ Column Densities}

To estimate the H$_2$ column densities of these {four}
clumps, we adopted the assumptions made in \cite{Li+etal+2007},
namely,

\begin{itemize}
  \item[(i)] The medium along a certain line of sight has a single
temperature (\citealt{Goldsmith+etal+1997});

\item[(ii)] The absorption coefficient at 350~$\upmu $m,
\textit{$Q$}(350), is $2\times $10$^{-4}$;

\item[(iii)] The characteristic grain radius is 0.1~$\upmu $m;

\item[(iv)] The grain density is 3~g~cm$^{-3}$;

\item[(v)] The gas to dust ratio (GDR) is 100.

\end{itemize}

The clump mass can be expressed as
\begin{eqnarray}\label{eq17}
M_{\rm{clump}}&=&0.10\,M_{\odot }\left [ \frac{2\times
10^{-4}}{Q\left (350 \right )} \right ]\left [\frac{\lambda
}{350~\upmu \rm{m}} \right ]^{3}
\nonumber\\
&& \left [ \frac{D}{1 \rm{kpc}} \right ]^{2}
 \left [ \frac{{\rm {GDR}}}{100} \right ]\left [ \frac{S\left (
\nu \right )}{\rm{Jy}} \right ]P_{\rm{f}}\left ( T_{\rm{d}} \right
),
\end{eqnarray}
where $S(\nu) $ is the flux density of the cloud at 350~$\upmu $m
at distance \textit{$D$}, in Jy, $T$$_{\rm{d}}$ is the dust
temperature, \textit{$P$}$_{\rm{f}}$(T$_{\rm{d}}$) is the Planck
factor and
\begin{equation}\label{eq18}
P_{\rm{f}}\left ( T_{\rm{d}} \right )=e^{h\nu /kT_{\rm{d}}}-1.
\end{equation}
In this formula (\citealt{Li+etal+2007}), the dust temperature is an
essential parameter. Although high density gases{ are} present in
these clumps, there can be {a }significant difference between the
dust and gas temperatures at the same position
(\citealt{Goldsmith+etal+1997}). However, when \textit{$n$}(H$_2$) =
10$^6$~cm$^{-3}$, the dust temperature approximately equals the gas
temperature (\citealt{Li+etal+2007}). Thus we can assume that along a
certain line of sight the dust temperature equals the local gas
temperature in this volume density condition. Since{ the} CS (5--4)
transition has a critical density of about 10$^6$~cm$^{-3}$, we
assume that within the boundaries of CS (5--4)-traced areas
\textit{$n$}(H$_2$) = 10$^6$~cm$^{-3}$. According to our non-LTE
analysis with RADEX (in Section 3.2.2), this assumption is reasonable
and we therefore adopt the approximation of dust temperature above in
the following estimate.

Here we assume  the gas temperature equals the kinetic temperature.
Based on the relation between the CS excitation temperatures and the
kinetic temperatures mentioned before, we actually adopted
 the calculated CS excitation temperatures as the
local gas temperatures and the dust temperatures along the same lines
of sight in the CS (5--4)-traced areas.

In the 350~$\upmu $m emission data, the grid spacing{ for the
sampling} is $1.5 \times1.5$~arcsec$^2$. By summing up the calculated
mass of every cell in the sampling grid, we calculated the total mass
of each clump\footnote{``The total mass'' is the sum of the
calculated mass of every single cell in each white box with solid
lines in Figure~\ref{G_350}. The boundaries of each box were
determined based on the CS{ }(5--4)-traced area, the area with
available calculated CS excitation temperatures and the profile of{
the} 350~$\upmu $m image (for G267.9--1.1).}. Subsequently we
estimated the H$_2$ column densities for each {of the }clumps cell by
cell~as
\begin{equation}\label{eq19}
N_{\rm{H_{2}(cell)}}=M_{\rm{clump(cell)}}\frac{f}{\left (D\theta \right )^{2}},
\end{equation}
where \textit{M}$_{\rm{clump(cell)}}$ is the calculated mass of every
single cell, $f$ is the mass fraction of H in gas with a $^4$He to H
ratio of 0.08459 (\citealt{Balser+2006}). \textit{$D$} is the
distance from the clump,{ and} $\theta$ is the sampling grid spacing,
which is 1.5$''$. Since the CS (5--4) line has a high critical
density {of} about 10$^6$~cm$^{-3}$ and referencing our non-LTE
analysis with RADEX (in Section 3.2.2) we assume that the element{ H}
is all in the form of H$_2$ in the dense gases within the CS
(5--4)-traced areas when {we }calculated the molecular hydrogen
column densities. The calculated clump masses (of the CS
(5--4)-traced areas, within the white boxes (solid lines) in
Fig.~\ref{G_350}) are listed in Column (6) of Table~\ref{complex}.
The average H$_2$ column densities in the areas {used to estimate
the} ortho-{\rm{H$_2$O}} abundance (white boxes (dashed lines) in
Fig.~\ref{G_350}) are listed in Column (9) of Table~\ref{complex}.
The H$_2$ column densities in these {four} clumps are {on} the order
of magnitude 10$^{22}$~cm$^{-2}$ or 10$^{23}$~cm$^{-2}$.

\subsection{Ortho-{\rm{H$_2$O}} Abundance}
\subsubsection{Method}

The observation of{ the} 557~GHz ortho-{\rm{H$_2$O}} 1$_{10}$ --
1$_{01}$ line was performed with SWAS with a pixel size of about $190
\times 190$arcsec$^2$ and a main beam efficiency of about 0.90
(\citealt{Melnick+1995}). This ground-state transition has{ a} large
spontaneous emission rate, $A$. As the stimulated absorption
coefficient $B$ is proportional to the spontaneous emission rate, it
leads to large opacities and makes excitation by photon trapping
important (\citealt{Wannier+etal+1991}). Since the collisional
de-excitation   rate coefficient $C$ is far less than $A$, this line
has a high critical density and the excitation is subthermal (in
other words, the de-excitation of upper level molecules is dominated
by emission photons rather than collisional de-excitation).  Thus,
although the line is expected to be optically thick at the line
center even for a relatively low water abundance, every collisionally
excited upper level molecule can always produce a photon which
finally escapes the cloud (\citealt{Snell+etal+2000a}). Thus, the
optically thick gas can be effectively thin
(\citealt{Snell+etal+2000a}). Therefore, the integrated antenna
temperature is proportional to the column density of
ortho-{\rm{H$_2$O}} under known temperature and H$_2$ volume density,
according to \cite{Snell+etal+2000a},
\begin{equation}\label{eq20}
\int T_{\rm{b}}dv  = Cn_{\rm{H_{2}}}\frac{c^{3}}{2\nu ^{3}k}N
\left ( {\rm{o-H_2O}} \right )\frac{h\nu }{4\pi }\exp
\left (\frac{-h\nu }{kT_{\rm{k}}} \right ).
\end{equation}
$T$$_{\rm{b}}$$dv$ is the integrated intensity, in K~km~s$^{-1}$.
$C$ is the collisional
  de-excitation rate coefficient from level 1$_{10}$ to level
1$_{01}$.

\subsubsection{Kinetic temperature and other details}

Equation~(\ref{eq20}) can be written as
\begin{equation}\label{eq21}
X\cdot \left( {\rm{o-H_2O}} \right)= a\cdot \frac{\int
T_{\rm{b}}dv }{N\left
 ( H_{2} \right )n_{\rm{H_{2}}}}\,,
\end{equation}
with
\begin{equation}\label{eq22}
a = \frac{1}{C\cdot \frac{c^{3}}{2\nu ^{3}k}\cdot \frac{h\nu
}{4\pi } \cdot \exp\left (\frac{-h\nu }{kT_{\rm{k}}}  \right )},
\end{equation}
where $a$ is a constant at a given temperature.

We adopted a{n} H$_2$ volume density of 10$^6$~cm$^{-3}$ in the
estimates since the areas where we estimated the ortho-{\rm{H$_2$O}}
abundances (the white boxes with dashed lines in Fig.~\ref{G_350})
are in the CS{ }(5--4)-traced areas. The average H$_2$ column
densities in the white boxes with dashed lines are listed in Column
(9) of Table~\ref{complex}.

The value of coefficient $a$ depends on the kinetic temperature
and the corresponding $C$. Assuming that CS lines and{ the}
ortho-{\rm{H$_2$O}} 1$_{10}$ -- 1$_{01}$ line originate  from the
same gas, we {take} the calculated CS excitation temperatures as
the kinetic temperature $T_{\rm{k}}$ at corresponding areas and
adopt them in the estimate of ortho-water abundance. Since the
pixel size of ortho-{\rm{H$_2$O}} data is much larger than that of
the CS lines (the sampling spacing), there is an average effect
for the kinetic temperature. We {calculated} the average
$T_{\rm{k}}$ and the corresponding standard deviation
(Table~\ref{complex}, Column (7) and (8)) to restrict the
temperature range for{ the} estimate. The collisional
de-excitation rate coefficients were calculated according to the
effective collisional excitation rate of ortho-{\rm{H$_2$O}} from
level 1$_{10}$ to level 1$_{01}$ (hereafter the effective
excitation rate) by para- or ortho-H$_2$ and the ortho{ }to{ }para
ratio of H$_2$. The effective excitation rates were adopted from
\cite{Dubernet+etal+2009} and
\cite{Daniel+etal+2011}\footnote{i.e., the ``effective rate
coefficient'' in these two papers} from 5\,K to 80\,K. Assuming
that H$_2$ molecules are in LTE, the ortho{ }to{ }para ratios of
H$_2$ were derived according to the H$_2$ rotational energy levels
from \cite{Dabrowski+1984} and the fractional population in the
H$_2$ rotational levels (\citealt{Phillips+etal+1996}).

\begin{table*}
\centering

\begin{minipage}{58mm}

\caption{The Parameters of the Four Clumps \label{complex}}
\end{minipage}
\fns \tabcolsep 1mm
 \begin{tabular}{lcccccccc}
  \hline\noalign{\smallskip}
Source     & RA J2000    & Dec J2000   & Distance  & Average \textit{$T$}$_{\rm{ex}}$ & Mass  &Average $T_{\rm{k}}$ & SD$^f$ of $T_{\rm{k}}$ & Average $N$(H$_2$) \\
 & & &(kpc) & (K)& ($M_{\odot }$) &  (K) & (K) & (cm$^{-2}$) \\
(1)& (2)& (3)& (4)& (5)& (6)& (7)& (8)& (9)\\
  \hline\noalign{\smallskip}
G267.9--1.1 & 08:59:12.00 & --47:29:04.0 & 1.5$^a$ & 26.5                                & $2.5\times 10^{3}$  &  \ \ 31.1  \ \ & 30.7\  & \  \ $3.7\times 10^{22}$  \\
G268.4--0.9 & 09:01:54.30 & --47:43:59.0 & 1.3$^b$ & 11.5                                & $4.6\times 10^{3}$  &  \ \ 11.5  \ \ & 1.4 \  & \  \ $1.2\times 10^{23}$  \\
G333.1--0.4 & 16:21:02.10 & --50:35:15.0 & 3.6$^c$ & --$^e$
                        & $1.2\times 10^{4}$  &  \ \ 31.9  \ \ & --
                          \  & \  \ $4.4\times 10^{22}$  \\
G336.5--1.5 & 16:40:00.20 & --48:51:20.0 & 1.4$^d$ & 16.2
                             & $9.0\times 10^{3}$
                             &  \ \ 17.8  \ \ & 6.3 \  & \  \ $8.5\times 10^{22}$  \\
  \hline\noalign{\smallskip}
\end{tabular}
\parbox{135mm}{Notes:
$^a$ \citealt{Frogel+Persson+1974}.
$^b$ \citealt{Zinchenko+etal+1995}.
$^c$ \citealt{Lockman+1979}.
$^d$ \citealt{Thompson+etal+2004}.
$^e$ We {could not} perform the estimate of CS
excitation temperatures for clump G333.1--0.4 since there {were} only
CS{ }(5{--}4) data.
We adopted its kinetic temperature (31.9~K) from
\cite{Lowe+etal+2014} and applied it in the following estimates.
This kinetic
temperature was derived from
the rotation temperature of the NH$_3$ (1,1) and (2,2)
transitions.
$^f$ ``SD'' stands for ``standard deviation.'' }

\end{table*}

\subsubsection{Ortho-{\rm{H$_2$O}} abundances}

The values of $a$ (see Eq.~(\ref{eq21})) and the estimated
ortho-{\rm{H$_2$O}} abundances at the kinetic temperatures for
every clump are listed in Table~\ref{New_abund}.

\begin{table*}
\centering

\begin{minipage}{85mm}

\caption{The Values of coefficient $a$ and Ortho-{\rm{H$_2$O}}
Abundances \label{New_abund}}\end{minipage}

\fns \tabcolsep 3mm
 \begin{tabular}{lccccc}
  \hline\noalign{\smallskip}
$T_{\rm{k}}$ (K) & \  \   $a$ \  \      & G267.9--1.1\         &
G268.4--0.9
           & G333.1--0.4           & G336.5--1.5 \\
         &  &  ($\times 10^{-10}$) & ($\times 10^{-10}$)&($\times 10^{-10}$)
           &($\times 10^{-10}$)\\
  \hline\noalign{\smallskip}
\  \ 5  \    \     & $3.7\times 10^{21}$  &$7.1
$  & --
               & --                    & -                     \\
\  \ 10 \    \     & $2.6\times 10^{20}$  &$7.0
$ &
$4.1
$ & --                    & $5.7
$  \\
\  \ 15 \    \     & $1.1\times 10^{20}$  &$7.1
$  &
$4.2
$ & --                    & $5.7
$  \\
\  \ 20 \    \     & $6.7\times 10^{19}$  &$7.0
$  & --
      & $7.2 
      $
       & $5.7 
       $
         \\
\  \ 30 \    \     & $3.6 \times 10^{19}$  &$5.8
$
  & --                    & $5.9
  $ & $4.7
  $  \\
\  \ 40 \    \     & $1.8\times 10^{19}$  &$3.7
$  & --                    & $3.8 
$ & --                     \\
\  \ 50 \    \     & $1.0\times 10^{19}$  &$2.4
$  & --                    & --                    & --                     \\
\  \ 60 \    \     & $7.2\times 10^{18}$  &$1.8
$  & --                    & --                    & --                     \\
\  \ 80 \    \     & $4.7\times 10^{18}$  &$1.3
$  & --                    & --                    & --                     \\
  \hline\noalign{\smallskip}
\end{tabular}
\end{table*}

The ortho-{\rm{H$_2$O}} abundances in the most {probable} temperature
ranges of these four clumps (i.e., G267.9-1.1, 30--40\ K;
G268.4--0.9, 10--15~K; G333.1--0.4, 30--40~K; G336.5--1.5, 15--20~K.
The corresponding ortho-{\rm{H$_2$O}} abundances were called ``the
typical ortho-{\rm{H$_2$O}} abundances'' in Section 4.1) are
presented in Figure~\ref{h2o_abun} together with some other
ortho-{\rm{H$_2$O}} water abundances of giant molecular cloud (GMC)
cores (\citealt{Snell+etal+2000b}) and molecular outflows
(\citealt{Franklin+etal+2008}), which {are} based on the same
ortho-{\rm{H$_2$O}} transition observed by SWAS. The
ortho-{\rm{H$_2$O}} abundances of these four clumps {are} at a low
level compared with other results.

\begin{figure*}[htbp]
\centering

\includegraphics[width=9.4cm]{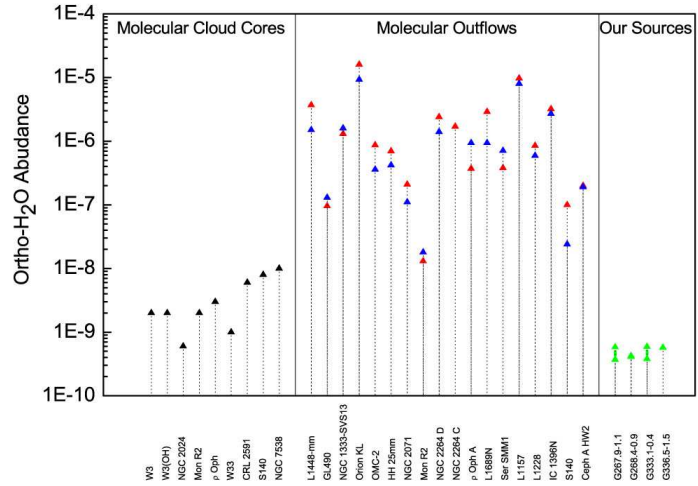}

 \caption{\baselineskip 3.6mm  The estimated ortho-{\rm{H$_2$O}}
abundances of clump G267.9--1.1, G268.4--0.9, G333.1--0.4 and
G336.5--1.5 in the most {probable} temperature ranges (G267.9--1.1,
30--40~K; G268.4--0.9, 10--15~K; G333.1--0.4, 30--40~K; G336.5--1.5,
15--20~K), in comparison with other results derived from{ the}
557~GHz ortho-{\rm{H$_2$O}} 1$_{10}$ -- 1$_{01}$ line observed by
SWAS. The solid black triangles represent the molecular cloud cores
from \cite{Snell+etal+2000b}. The solid red triangles and solid blue
triangles represent the redshifted and blueshifted emission
respectively in each molecular outflow from
\cite{Franklin+etal+2008}. The solid green triangles linked with
green lines show the ortho-{\rm{H$_2$O}} abundance ranges of the four
clumps in this paper.}
 \label{h2o_abun}
\end{figure*}

\subsection{{\rm{N$_2$H$^+$}} Abundances}

Since the critical density of{ the} {\rm{N$_2$H$^+$}} (1--0) line is far
lower than that of the CS (5--4) line, we can assume that CS lines
((5--4) and (2--1)) and{ the} {\rm{N$_2$H$^+$}} (1--0) line are all
thermally populated in the CS (5--4)-traced areas, thus their
excitation temperatures{ are} all approximately equal to the kinetic
temperatures. In this situation, we can adopt the excitation
temperatures of CS as the excitation temperatures of
{\rm{N$_2$H$^+$}} at the same positions. Then with the same
assumptions and approximations we used in calculation of the
column density of upper level CS molecules, we calculate the
corrected {\rm{N$_2$H$^+$}} column density at $J = 1$. The
{\rm{N$_2$H$^+$}} column density $N_{\rm{total}}$ is estimated as
\begin{equation}\label{eq23}
N_{\rm{total}}=N_{\rm{J=1}}\frac{Z}{2J+1}\exp \left [
\frac{hB_{\rm{e}}J(J+1)}{kT} \right ]\,,
\end{equation}
according to \cite{Rohlfs+Wilson+1996}. \textit{$N$}$_{\rm{J=1}}$
is the {\rm{N$_2$H$^+$}} column density without the Rayleigh-Jeans
approximation at $J = 1$. $B_{\rm{e}}$ is the rotational constant
of{ the} {\rm{N$_2$H$^+$}} molecule at vibrational energy level $v
= 0$ \footnote{\it
http://www.cv.nrao.edu/php/splat/species\_metadata\_displayer.php?
species\_id=148}.
$J$ is the rotational quantum number of{ the} upper level and $J =
1$. $k$ is the Boltzmann constant and $h$ is the Planck constant.
We adopt the excitation temperatures of CS as the temperature $T$.
$Z$ is the rotational partition function of {\rm{N$_2$H$^+$}}.
Since {\rm{N$_2$H$^+$}} is {a} linear molecule
(\citealt{Mangum+Shirley+2015}), when the contribution of the
vibrational excited states {is} not take{n} into account,
\begin{equation}\label{eq24}
Z \simeq  \sum_{J=0}^{\infty }\left ( 2J+1 \right )\exp\left
( -\frac{hB_{0}J\left ( J+1 \right )}{kT} \right ){\,.}
\end{equation}
$B_{0}$ is the rigid rotor rotational constant of{ the}
{\rm{N$_2$H$^+$}} molecule at the ground vibrational state $v = 0$
and $B_{0} = 46586.88$~MHz (\citealt{Mangum+Shirley+2015}). $J$
 is
the rotational quantum number and $J = 0, 1, 2, \ldots, k$. $h$
and $T$ are as the same as in Equation~(\ref{eq23}).

According to \cite{Mangum+Shirley+2015}, if we use one or several
hyperfine transition(s) that can be observed to derive the column
density of N$_2$H$^+$, we must take the relative line strengths of
the hyperfine transition(s) into consideration. However, in our
observation the spectra cover all the {seven} hyperfine
transitions which can be observed in the $J = 1 \to  0$
transition. Thus, we just calculate the rotational partition
functions at corresponding temperatures and then estimate the
{\rm{N$_2$H$^+$}} column densities. We average the estimated H$_2$
column densities at every {\rm{N$_2$H$^+$}} pixel and then
estimate the {\rm{N$_2$H$^+$}} abundances. The results are {shown}
in Table~\ref{n2hp_abun}  and Figure~\ref{n2hp_abun_map} and all
offsets are relative to the corresponding coordinates (J2000) in
Table~\ref{complex} and the unit is arcsec.

\begin{table*}
\centering

\begin{minipage}{40mm}

\caption{{\rm{N$_2$H$^+$}} Abundances \label{n2hp_abun}}
\end{minipage}

\fns \tabcolsep 10mm
 \begin{tabular}{ll}
  \hline\noalign{\smallskip}
Clump      & {\rm{N$_2$H$^+$}} Abundances  \\
  \hline\noalign{\smallskip}
G267.9--1.1 & $1.0\times 10^{-10}$ -- $1.5\times 10^{-8}$ \\
G268.4--0.9 & $6.1\times 10^{-11}$ -- $4.3\times 10^{-9}$ \\
G333.1--0.4 & $2.6\times 10^{-10}$ -- $4.2\times 10^{-9}$ \\
G336.5--1.5 & $5.6\times 10^{-11}$ -- $1.4\times 10^{-9}$ \\
  \hline\noalign{\smallskip}
\end{tabular}
\end{table*}

\begin{figure*}
\centering

\vs
\includegraphics[width=9.5cm]{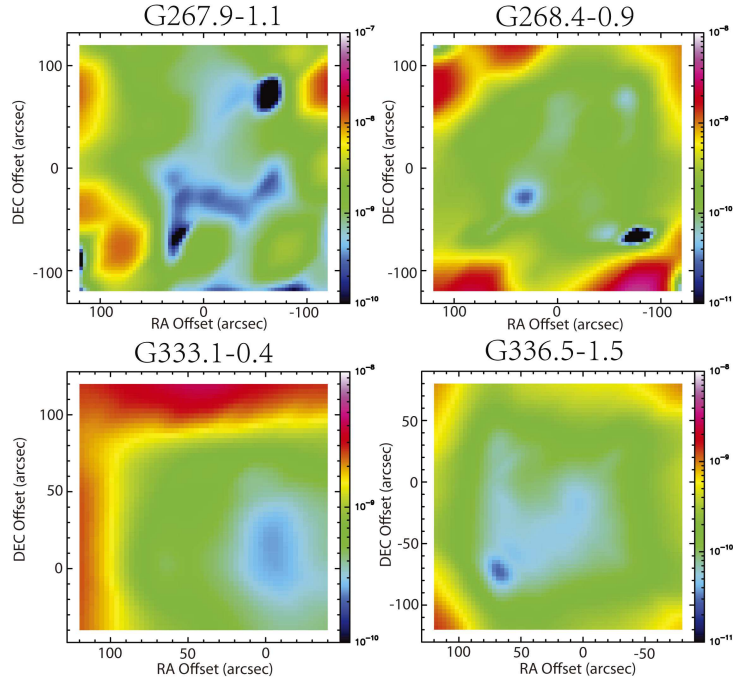}

\caption{\baselineskip 3.6mm  {\it{Upper left}}: G267.9--1.1 {\rm
N$_2$H$^+$} abundance. {\it{Upper right}}: G268.4--0.9 {\rm
N$_2$H$^+$} abundance. {\it{Lower left}}: G333.1--0.4 {\rm
N$_2$H$^+$} abundance. {\it{Lower right}}: G336.5--1.5 {\rm
N$_2$H$^+$} abundance. The color bars are abundance scales.}
 \label{n2hp_abun_map}
\end{figure*}

\section{Discussion}
\subsection{Ortho-{\rm{H$_2$O}} Abundances}

The typical ortho-{\rm{H$_2$O}} abundances of these
{four} clumps are in the range $3.7\times 10^{-10}$ --
$5.8\times 10^{-10}$ for G267.9--1.1, $4.1\times 10^{-10}$ --
$4.2\times 10^{-10}$ for G268.4--0.9, $3.8\times 10^{-10}$ --
{$5.9\times 10^{-10}$} for
G333.1--0.4 and around $5.7\times 10^{-10}$ for G336.5--1.5.

The typical ortho-{\rm{H$_2$O}} abundances are at a low level
compared with those of cold ($T <50$~K) {GMC} cores estimated with
the same principle by \cite{Snell+etal+2000b}. The upper limits of
the ortho-{\rm{H$_2$O}} abundances of these four clumps are {on} the
order of $10^{-10}$, lower than the abundances of most of the {GMC}
cores in \cite{Snell+etal+2000b}. In our estimate, the effective
excitation rates we adopted for para-H$_2$ (${j} = 0$) {are} larger
than those from \cite{Phillips+etal+1996} by a factor of $\sim$1--3
at temperatures from 20\,K to 80\,K. This fact should be noticed when
comparing our results with the results in \cite{Snell+etal+2000b} or
\cite{Franklin+etal+2008} (see Fig.~\ref{h2o_abun}).

The low abundances may be caused by the low temperatures of these
clumps if we consider the water vapor originating from the interior
of the clumps. Even at temperatures as low as 10\,K, water can form
in the ISM (\citealt{vanDishoeck+etal+2013}). However, at such low
temperatures and high densities, the freeze-out procedure dominates
(\citealt{Bergin+vanDishoeck+2012}) and until the temperature is
above about 100\,K (\citealt{Hollenbach+etal+2009}), water molecules
can be desorbed through thermal sublimation. Also, in the interior of
the dense clump, the desorption of frozen water molecules is unlikely
to be caused by photodesorption. Although there is an average effect
for the temperature in the large SWAS beam (and in the
ortho-{\rm{H$_2$O}} pixels, also), we can find that the areas along
the line of sight from where the ortho-{\rm{H$_2$O}} emissions
origin{ate} are quite cold, or in other words, {are }not warm enough
to produce much water vapor. Consequently the ortho-{\rm{H$_2$O}}
abundances are low.

On the other hand, since these clumps {are }all locate{d} in star
forming regions, the ortho-{\rm{H$_2$O}} emission therefore can
originate {primarily} at the intermediate depth of these clumps
where neither the photodissociation nor the freezing out of the
\rm{H$_2$O} molecules, but the photodesorption process dominates
according to a model for the temperature and chemical structure in
molecular clouds (\citealt{Hollenbach+etal+2009}). Thus, {since} we
took the H$_2$ column densities along the line of sight to estimate
the abundances, it consequently results in apparent low
ortho-{\rm{H$_2$O}} abundance for the whole clump while in the
photodesorbed layer the water vapor is actually more abundant.

{In addition}, there are also considerations {of} possible fact{or}s
which can cause the apparent low ortho-{\rm{H$_2$O}} abundances of
those clumps but have been masked due to the averaging effect of the
large SWAS beam. For example, small structures in the clump such as a
hot outflow may contribute the majority of the observed gaseous
water. The water abundance may {vary greatly} within the same clump.
{A s}imilar phenomenon has been confirmed in the outflow powered by
L1157-mm (a low-mass Class 0 protostar), {in which} the water
abundance of {the }hot component is about two orders of magnitude
higher than that of the nearby colder component
(\citealt{Busquet+etal+2014}). In the areas we estimated ortho-water
abundance, the maximum CS excitation temperatures of G267.9--1.1,
G268.4--0.9 and G336.5--1.5 are 165.2\,K (and the second largest
value is 86.6\,K), 15.8\,K and 42.3\,K, respectively (and for
G333.1--0.4, the average kinetic temperature is 31.9\,K, according to
\cite{Lowe+etal+2014}). For G267.9--1.1, there is {a }possibility
that the warm component makes a big contribution to the origination
of gaseous water. However, we {cannot} infer more information on
structures smaller than the SWAS beam size which can further reveal
the origination of gaseous water.

If{ the} ortho-{\rm{H$_2$O}} 1$_{10}$ -- 1$_{01}$ line originates
from the same gas as CS lines, as we assumed when we estimated the
ortho-water abundance, then the CS lines may help to find traces of
outflows. The CS spectra of G267.9--1.1 have b{r}oad wings. We found
that 267.9--1.1 as well as G268.4--0.9 does show velocity variation
over the clump on its channel map, but the spatial resolution of CS
data is not high enough to identify the outflows.
\cite{Lapinov+etal+1998} {have} mapped G268.4--0.9 (G268.42--0.85) in
CS $J = 5 - 4$ (SEST, $20''$ sampling spacing) and $J = 7 - 6$ ({t}he
CSO telescope, $10''$ sampling spacing) lines. They used {the
}Maximum Entropy Method{ (MEM}) deconvolution technique to {achieve}
higher angular resolution (\citealt{Lapinov+etal+1998}). In their
study, the CS (5--4) map shows two peaks with {an} LSR velocity
difference of about 0.7~Km~s$^{-1}$ and on the CS (7--6) map, a
bipolar structure was identified but no further temperature
information was offered for this bipolar structure
(\citealt{Lapinov+etal+1998}).

To be honest and objective, it is kind of arbitrary to assume such
{a} high density over the whole area in which we estimated the
ortho-{\rm{H$_2$O}} abundance in each clump. There are {very} likely
to be H$_2$ density gradients in these areas. If we adopt
10$^4$~cm$^{-3}$ rather than 10$^6$~cm$^{-3}$ as the H$_2$ density in
the ortho-{\rm{H$_2$O}} abundance estimation, then the typical
ortho-{\rm{H$_2$O}} abundances will be at the magnitude of 10$^{-8}$,
the same as those of most of the {GMC} cores in
\cite{Snell+etal+2000b}.

\subsection{{\rm{N$_2$H$^+$}} Abundances}

The {\rm{N$_2$H$^+$}} abundances of these {four} clumps
are in the range of $1.0\times 10^{-10}$ - $1.5\times 10^{-8}$ for
G267.9--1.1, $6.1\times 10^{-11}$ - $4.3\times 10^{-9}$ for
G268.4--0.9, $2.6\times 10^{-10}$ - $4.2\times 10^{-9}$ for
G333.1--0.4 and $5.6\times 10^{-11}$ - $1.4\times 10^{-9}$ for
G336.5--1.5. The distribution of {\rm{N$_2$H$^+$}} abundance in each
clump has a common decreasing trend toward the center. Although the
abundance distributions we {derived} are only projected results {in} a plane
perpendicular to the line of sight, we noticed that
\cite{Melnick+etal+2011} suggested that {\rm{N$_2$H$^+$}} {is} likely
to{ be} distribute{d} {primarily} in the clump rather than
in the surface layers. When it comes to the depletion of
{\rm{N$_2$H$^+$}}, CO and electrons are the major destroyers of
{\rm{N$_2$H$^+$} in the gas phase and their reaction with
{\rm{N$_2$H$^+$} generates N$_2$ (\citealt{Aikawa+etal+2001},
\citealt{Aikawa+etal+2005}). According to
\cite{Bergin+Tafalla+2007}, in the dense cores the neutrals
(including CO) will{ rapidly} freeze onto the grains. Consequently,
the abundance of {\rm{N$_2$H$^+$} will increase as a result of the
disappearance of CO (\citealt{Bergin+Tafalla+2007}). However, when
we focus on temperature, we notice that when {the }dust
temperature rises from 10\,K to about 30\,K, CO begins to sublimate
(\citealt{vanDishoeck+etal+2013}). Thus, based on the gas
temperatures we estimated ({t}he dust temperatures
{are }approximately equal to local gas temperatures at
such a high density as we had assumed), we can infer that in
the high density center of the clump, the gas and dust are warm
enough and the gaseous CO is abundant. {\rm{N$_2$H$^+$} is therefore
depleted by CO and that leads to a drop {in} {\rm{N$_2$H$^+$}}
abundance toward the center.

\section{conclusions}

We studied G267.9--1.1, G268.4--0.9, G333.1--0.4 and G336.5--1.5,
four of the brightest ortho-{\rm{H$_2$O}}
 sources in the southern
sky observed by SWAS. We estimated their CS excitation temperatures in
the CS (5--4)-traced areas and estimated their masses. Based on the
temperatures and the masses, we then derived their average H$_2$
column densities and estimated the ortho-{\rm{H$_2$O}} and
{\rm{N$_2$H$^+$}} abundances.

The typical molecular clumps in our study have H$_2$ column densities
of $\sim 10 ^{22}$ to $10 ^{23}$~cm$^{-2}$ and ortho-{\rm{H$_2$O}}
abundances of 10$^{-10}$. The low ortho-{\rm{H$_2$O}} abundances can
be caused by the freeze-out of {\rm{H$_2$O}} in the interior of the
clumps due to the low temperatures if the ortho-{\rm{H$_2$O}}
originates from the interior of the clumps.

The typical {\rm{N$_2$H$^+$}} abundances of these four clumps in
this paper range from 10$^{-11}$ to 10$^{-9}$. Since in the center
areas of the clumps, dust at such a high density {is}
at temperatures{ such} that CO can be released into the gas phase, the
common trend of abundance decreasing toward the center of the clump
can be a result of the depletion of {\rm{N$_2$H$^+$}} caused by CO.

\begin{acknowledgements}
We sincerely thank the anonymous referee and the scientific editor
for their wholehearted and patient help and the valuable advice they
provided to help us improve this paper. This research is supported by
{the }National Basic Research Program of China {(973 program, Nos.
2012CB821800 and 2015CB857100)}, {the }National Natural Science
Foundation of China {(}No. 11373038{)} and the Strategic Priority
Research Program ``The Emergence of Cosmological Structures" of the
Chinese Academy of Sciences {(}Grant No. XDB09000000{)}.
\end{acknowledgements}

\appendix
\section{Supplementary Material}

In the temperature range (20\,K to 80\,K), para- and ortho-H$_2$ are
barely populated at energy levels other than $j_2$ = 0 and $j_2$ = 1
($j_2$ is the rotational level of H$_2$), respectively. We adopted
the effective rate coefficients at $j_2$ = 0 and $j_2$ = 1 from
\cite{Dubernet+etal+2009} and \cite{Daniel+etal+2011}.

For G267.9--1.1 and G336.5--1.5, we estimated the ortho-{\rm{H$_2$O}}
abundance at the ortho-{\rm{H$_2$O}} integrated intensity maximum
pixel in FITS format   data.  Although there is more than one
sampling cell having emission in the CLASS format data, when written
in FITS format,  only the pixel with maximum integrated intensity
(190$\times$190~arcsec$^2$, the boxes with white dashed lines in
Fig.~\ref{G_350}) are located within the areas with calculated H$_2$
column densities. However, the estimated ortho-{\rm{H$_2$O}}
abundance of the pixels with maximum integrated intensity still
characterize the ortho-{\rm{H$_2$O}} abundance of these clump in a
sense.  For G268.4--0.9 and G333.1--0.4, the CLASS format
ortho-{\rm{H$_2$O}} data only have one sampling cell and the center
of the cell is the same as the corresponding coordinates in Table~3
(with a deviation of 0.01\,s on RA). So we just estimated the
ortho-{\rm{H$_2$O}} abundance in the only sampling cell (the box with
white dashed lines in Fig.~\ref{G_350}, 190$\times$190~arcsec$^2$,
the same as the pixel size of the FITS format data).

We calculate the ortho-{\rm{H$_2$O}} abundance of G336.5--1.5 in
the overlap{ping} area of the white box (dashed lines) and the white box
(solid lines) with corresponding average $T_{\rm{k}}$, average
H$_2$ column density and proportionally corrected
ortho-{\rm{H$_2$O}} integrated intensity.

When calculating the partition functions of CS and N$_2$H$^+$, we
summed the polynomial term by term from the $J = 0$ level, following
the incremental rotational quantum number. When the value of a term
is less than 0.1\% of the sum of all the terms of the lower levels,
then this term will be the last term added in the summation.

\end{document}